\documentclass[12pt]{article}

\usepackage{amssymb}
\usepackage{amsmath}
\usepackage{amscd}
\usepackage{latexsym}
\usepackage{graphicx}

\usepackage{cite}

\newcommand{\be}{\begin{equation}}
\newcommand{\ee}{\end{equation}}
\newcommand{\Dlt}{\Delta}
\newcommand{\dlt}{\delta}
\newcommand{\prt}{\partial}
\newcommand{\br}{{\bf r}}
\newcommand{\ba}{{\bf a}}
\newcommand{\bk}{{\bf k}}
\newcommand{\bfe}{{\bf e}}

\newcommand{\bS}{{\bf S}}
\newcommand{\bp}{{\bf p}}
\newcommand{\bq}{{\bf q}}
\newcommand{\bu}{{\bf u}}

\newcommand{\bt}{\beta}
\newcommand{\vp}{\varphi}
\newcommand{\ep}{\varepsilon}
\newcommand{\al}{\alpha}
\newcommand{\ra}{\rightarrow}
\newcommand{\sgm}{\sigma}

\newcommand{\gm}{\gamma}
\newcommand{\om}{\omega}
\newcommand{\Om}{\Omega}
\newcommand{\Gm}{\Gamma}
\newcommand{\dgr}{\dagger}
\newcommand{\lbd}{\lambda}
\newcommand{\Lbd}{\Lambda}
\newcommand{\rgl}{\rangle}
\newcommand{\lgl}{\langle}
\newcommand{\cH}{{\cal H}}
\newcommand{\cL}{{\cal L}}
\newcommand{\cA}{{\cal A}}
\newcommand{\cF}{{\cal F}}
\newcommand{\cD}{{\cal D}}

\begin{document}

\begin{center}
 
{\Large{\bf Mesoscopic Fluctuations in Statistical Systems} \\ [5mm]

V.I. Yukalov$^{1,2}$ and E.P. Yukalova$^{3}$}  \\ [3mm]

{\it
$^1$Bogolubov Laboratory of Theoretical Physics, \\
Joint Institute for Nuclear Research, Dubna 141980, Russia \\ [2mm]

$^2$Instituto de Fisica de S\~ao Carlos, Universidade de S\~ao Paulo, \\
CP 369, S\~ao Carlos 13560-970, S\~ao Paulo, Brazil \\ [2mm]

$^3$Laboratory of Information Technologies, \\
Joint Institute for Nuclear Research, Dubna 141980, Russia } \\ [3mm]

{\bf E-mails}: {\it yukalov@theor.jinr.ru}, ~~ {\it yukalova@theor.jinr.ru} \\

\end{center}

\vskip 1cm

\begin{abstract}

The fluctuations are termed mesoscopic, when their typical size is essentially larger 
then the average distance between the nearest neighbors, while being much smaller than 
the overall system size. Since the features of mesoscopic fluctuations are essentially 
different from those of the surrounding matter, they can be interpreted as fluctuations 
of one phase occurring inside another host phase. In condensed matter, these fluctuations 
are of nanosize. They can occur in many-body systems of different nature, for instance, 
they are typical for condensed matter, can appear in systems of trapped atoms, and also 
arise in biological and social systems. A survey of the experimental evidence for the 
occurrence of mesoscopic fluctuations in different materials and systems is given. The 
main attention is paid to a general theoretical approach for describing them. Applications 
of the approach are also discussed.          

\end{abstract}

\vskip 2cm
{\parindent=0pt
{\it Keywords}: mesoscopic fluctuations, condensed matter, trapped atoms, phase 
transitions    }

\newpage

\section*{Contents}
   
{\parindent=0pt
{\bf 1}. Introduction

\vskip 3mm
{\bf 2}. Examples of mesoscopic fluctuations

\vskip 2mm
\hspace{0.4cm}   2.1. Crystal-liquid phase transition
\vskip 2mm
\hspace{0.4cm}   2.2. Structural phase transformations
\vskip 2mm
\hspace{0.4cm}   2.3. Coexistence of magnetic and nonmagnetic phases
\vskip 2mm
\hspace{0.4cm}   2.4. Coexistence of ferroelectric and paraelectric phases
\vskip 2mm
\hspace{0.4cm}   2.5. Colossal magnetoresistance materials
\vskip 2mm
\hspace{0.4cm}   2.6. High-temperature superconductors
\vskip 2mm
\hspace{0.4cm}   2.7. Trapped Bose-condensed systems
\vskip 2mm
\hspace{0.4cm}   2.8. Fluctuating social groups

\vskip 3mm
{\bf 3}. Spatial and temporal scales

\vskip 3mm
{\bf 4}. Theory of mesoscopic fluctuations

\vskip 2mm
\hspace{0.4cm}   4.1. Quantum statistical ensemble
\vskip 2mm
\hspace{0.4cm}   4.2 Methods of phase selection
\vskip 2mm
\hspace{0.4cm}   4.3. Weighted Hilbert spaces
\vskip 2mm
\hspace{0.4cm}   4.4. Spatial phase configuration
\vskip 2mm
\hspace{0.4cm}   4.5. Operators of local observables
\vskip 2mm
\hspace{0.4cm}   4.6. Heterophase statistical operator
\vskip 2mm
\hspace{0.4cm}   4.7. Averaging over phase configurations
\vskip 2mm
\hspace{0.4cm}   4.8. Surface thermodynamic potential
\vskip 2mm
\hspace{0.4cm}   4.9. Thermodynamics of heterophase systems

\vskip 3mm
{\bf 5}. Classical heterophase systems

\vskip 2mm
\hspace{0.4cm}   5.1. Classical statistical ensemble
\vskip 2mm
\hspace{0.4cm}   5.2. Selection of thermodynamic phases
\vskip 2mm
\hspace{0.4cm}   5.3. Classical heterophase probability distribution
\vskip 2mm
\hspace{0.4cm}   5.4. Results of averaging over phase configurations
\vskip 2mm
\hspace{0.4cm}   5.5. Quasiaverages and weighted spaces

\vskip 3mm
{\bf 6}. Models of heterophase systems

\vskip 2mm
\hspace{0.4cm}   6.1. Magnetic materials with paramagnetic fluctuations
\vskip 2mm
\hspace{0.4cm}   6.2. Ferroelectrics with paraelectric fluctuations
\vskip 2mm
\hspace{0.4cm}   6.3. Systems with mesoscopic density fluctuations
\vskip 2mm
\hspace{0.4cm}   6.4. Mesoscopic structural fluctuations
\vskip 2mm
\hspace{0.4cm}   6.5. Fluctuations in frustrated materials
\vskip 2mm
\hspace{0.4cm}   6.6. Superfluid fluctuations in solids
\vskip 2mm
\hspace{0.4cm}   6.7. Superconductors with mesoscopic fluctuations
\vskip 2mm
\hspace{0.4cm}   6.8. Debye-Waller and M\"{o}ssbauer factors

\vskip 3mm
{\bf 7}. Conclusion   }

\newpage

\section{Introduction}

The appearance of mesoscopic fluctuations is widely known to occur in a number 
of many-body systems, such as condensed-matter systems, systems of trapped atoms,
as well as in biological and social systems. Usually, the features of mesoscopic 
fluctuations essentially differ from those of the host phase, which allows for their 
interpretation as phase fluctuations. They can be called by different names, such 
as heterophase fluctuations, nanoscale phase separation, nanoscale phase fluctuations, 
and like that. These guises have the main common features corresponding to the 
appearance of local fluctuations of one phase inside another host material. These 
fluctuations usually are randomly distributed in space and in time, which makes their 
description highly complicated.   

Let us specify why the fluctuations are called mesoscopic. Generally, there are for 
this two reasons. First, the sizes of the phase embryos are intermediate between the
average distance from the nearest neighbors and the overall system size. Say, the 
average distance between the particles in a condensed-matter system, or between neurons 
in the brain, or between agents in a society is $a$, while the overall size of the 
considered system is $l_{exp}$. Then the typical size $l_{mes}$ of a mesoscopic phase 
embryo is in the range
\be
\label{1}
a \; \ll \; l_{mes} \; \ll \; l_{exp} \;   .
\ee

The other reason for calling these fluctuations mesoscopic is the duration of their 
lifetime $t_{mes}$ that usually is between the interaction time $t_{int}$ and the 
experimental observation time $t_{exp}$,
\be
\label{2}
 t_{int} \; \ll \; t_{mes} \; \ll \; t_{exp} \;   .
\ee
The fluctuation lifetime has to be larger than the interaction time in order that 
the embryo would really represent another phase, but not just a locally nonequilibrium 
microscopic fluctuation of the same phase.  

The mesoscopic phase fluctuations are usually randomly distributed in the space of 
the sample. The probability of their appearance is not prescribed but has to be 
self-consistently defined by the considered system properties. Strictly speaking, 
on microscopic scale, a system with mesoscopic fluctuations is not equilibrium but 
can be only quasi-equilibrium. However, on macroscopic scale, corresponding to a 
sample averaged over space and time, the renormalized system looks as equilibrium, 
which allows for the development of a quasi-equilibrium approach and the construction 
of effective statistical models. Of course, heterophase fluctuations can also appear
in strongly nonequilibrium systems, as will be considered below.  

The ideas on the occurrence of heterophase fluctuations in condensed matter have long
history, first being applied to water-ice phase transition. It has been conjectured 
\cite{Rontgen_1,Brody_2,Bernal_3} that below the melting temperature water contains
ice-like clusters, while above the melting point there arise local liquid-like
fluctuations in ice. Frenkel \cite{Frenkel_4,Frenkel_5,Frenkel_6} emphasized that 
heterophase fluctuations can appear in different types of condensed matter and their    
account is necessary for the accurate description of condensed-matter systems. There 
have been a number of experimental works confirming the existence of these fluctuations 
in condensed matter, with taking them into account by means of phenomenological models. 

The main characteristics of such mesoscopic fluctuations depend on the matter where they 
appear \cite{Khait_7,Yukalov_8,Yukalov_9}. For example, in condensed matter the typical 
length and time scales are as follows. The interaction radius is $r_{int} \sim 10^{-8}$ cm,
the radius of a mesoscopic fluctuation is of nanosize, being $l_{mes} \sim 10^{-7}$ cm, 
the interaction time is $t_{int} \sim 10^{-14}$ s, the local equilibration time is 
$t_{loc} \sim 10^{-13}$ s, and the mesoscopic fluctuation time is $t_{mes} \sim 10^{-12}$s
or longer. 

In trapped atomic clouds, the parameters can be widely varied. For instance, in a 
nonequilibrium droplet state \cite{Yukalov_10,Yukalov_11,Yukalov_12}, the typical 
interaction time can be $t_{int} \sim 10^{-8}$ s and the local equilibration time, 
$t_{loc} \sim 10^{-3}$ s, while the coherence length, defining the droplet size, 
$l_{coh} \sim 10^{-5}$ cm, and the fluctuating droplet lifetime, $t_{mes} \sim 10^{-2}$ s.  

In nuclear matter, the strong-interaction time is $t_{int} \sim 10^{-24}$ s and the local
equilibration time is $t_{loc} \sim 10^{-23}$ s. For realistic parameters, the deconfinement
transition is a crossover, where there can coexist droplets of hadron phase intermixed
with quark-gluon phase \cite{Yukalov_13,Yukalov_128}.
 
The specific features of the present review are as follows.
\vskip 2mm
(i) We give a survey of recent literature on the subject, emphasizing the points that, 
to our mind, require additional attention.

\vskip 2mm
(ii) We present a general statistical approach allowing for the description of different 
systems with mesoscopic phase fluctuations. This approach is applicable to classical as well 
as quantum systems.

\vskip 2mm
(iii) Using the developed general approach, we consider not only condensed matter but 
also discuss other systems with mesoscopic fluctuations, such as cold trapped atoms and 
biological or social systems.

\section{Examples of mesoscopic fluctuations}

Discussing the materials and systems, where mesoscopic phase fluctuations have been 
observed, we can mention only some of them, since it is impossible to list all voluminous 
literature. Where possible, we prefer to cite review type publications. Quite a number 
of references on heterophase fluctuations can be found in the earlier reviews 
\cite{Khait_7,Yukalov_8,Yukalov_9}. More information on mesoscopic fluctuations in 
particular materials will be cited in the appropriate places below. We shall try to cite 
the references that have not been mentioned in the previous reviews \cite{Yukalov_8,Yukalov_9}, 
unless these references are important. Our aim is not to present an exhaustive list of 
references but rather to give typical examples of mesoscopic heterophase fluctuations 
usually arising around phase transitions.

\subsection{Crystal-liquid phase transition}   

The understanding that in a crystal before its melting there appear fluctuational regions 
of disorder imitating liquid state, while above the melting point there survive 
quasi-crystalline clusters, has been accepted long time ago, as can be inferred from 
the review works \cite{Ubbelohde_14,Hayes_15,Dash_16}. The lifetime of these fluctuations
is estimated as $10^{-12}$ s and the typical diameter between $10^{-6}$ and $10^{-7}$ cm
\cite{Khait_17,Khait_18,Khait_19}. Such local mesoscopic fluctuations can arise in large
systems as well as in finite systems \cite{Wales_20}. Molecular dynamics calculations
\cite{Tanaka_21,Ohmine_22,Shiratani_23} confirm that in water and other liquids, in 
addition to the standard vibrational oscillations of order $10^{-14}-10^{-13}$ s, there 
exist large local energy fluctuations of order $10^{-12}-10^{-11}$ s, associated with 
the local structural changes, involving about $10-40$ molecules. The local density 
structures, leading to crystallization, are rather stable during their lifetime. These 
ice-type clusters are randomly distributed in a sea of normal disordered liquid 
\cite{Tanaka_24,Tanaka_25,Tanaka_26}. This phenomenon exists not only for water but
for many liquid states \cite{Pajak_27,Haskel_28,Szafranska_42}. Thus, generally, 
liquid is not a homogeneous substance and in any liquid there exist local structures 
arising as medium-scale fluctuations. Cooperative medium-range ordering is a universal 
feature of different liquids \cite{Tanaka_29}. Liquid-solid phase transition happens 
being triggered by the appearance of heterophase fluctuations.

It is necessary to emphasize that the temperature of melting is not a temperature of
mechanical instability that is unrelated to the melting transition. The models of pure 
solids, without heterophase fluctuations, do possess the points of mechanical instability, 
however these instability points are several times higher than the melting points,
as has been reliably proved in many calculations \cite{Zubov_30,Zubov_31,Cotterill_32,
Albers_33,Cotterill_34,Moleko_35,Yukalov_36}. 

Two-dimensional melting, according to the Monte Carlo simulations 
\cite{Abraham_37,Barker_38,Stranburg_39} and molecular dynamics calculations 
\cite{Weber_40}, is also a first-order liquid-solid phase transition, in the same way 
as the three-dimensional melting. The supposed hexatic phase, arising in some calculations, 
could be due to the influence of initial and boundary conditions \cite{Naidoo_41}.

\subsection{Structural phase transformations}

Mesoscopic phase fluctuations exist in the vicinity of many structural phase transitions, 
where two phases with different crystallographic structures coexist. This has been proved
by molecular dynamic simulations \cite{Morris_43}. The precursor effects are typical of
martensitic transformations \cite{Morris_44,Clapp_45}. For instance, they have been 
observed \cite{Przenioslo_46,Przenioslo_47} in CaMn$_7$O$_{12}$ by using high-resolution 
synchrotron and neutron powder diffraction. More examples can be found in 
\cite{Bruce_48,Yukalov_49}.

Strictly speaking, solid-liquid phase transition is also a kind of structural 
transformation, since the spatial structures of a uniform liquid and a crystallographic
lattice are different.

\subsection{Coexistence of magnetic and nonmagnetic phases}

Magnetic phases, such as ferromagnetic or antiferromagnetic phases, often coexist with 
nonmagnetic phases. Thus neutron diffraction, high-resolution electron microscopy, and
magnetic-susceptibility measurements with La$_{1-x}$Sr$_x$CoO$_3$ perovskites give direct 
evidence for the nanoscopic phase separation of the material into hole-rich ferromagnetic
regions and a hole-poor semiconducting matrix \cite{Cacuiffo_50}.  Perovskites clearly 
exhibit the existence of short-range ferromagnetic order above $T_c$ \cite{Cacuiffo_65}.
Using electron holography and Fresnel imaging of La$_{0.5}$Ca$_{0.5}$MnO$_3$, it was found 
that the material contains ferromagnetic regions coexisting with regions with no local 
magnetization \cite{Loudon_51}. Some binary alloys demonstrate coexistence of ferromagnetic 
and nonmagnetic local properties \cite{Endo_52}. Vast literature on the nanoscale coexistence 
of ferromagnetic and antiferromagnetic phases, ferromagnetic and paramagnetic phases, 
antiferromagnetic and paramagnetic phases, magnetic and spin-glass phases, and phases with 
different magnetic orientations is given in \cite{Yukalov_49}. Nanoscale phase separation 
into magnetic and nonmagnetic phases is a typical phenomenon for many semiconductors 
\cite{Nagaev_53,Nagaev_54,Nagaev_55,Kagan_56}. Several metals, even such as Fe and Ni,
show the occurrence of short-range ferromagnetic order above $T_c$, which leads to the 
existence of spin waves \cite{Lynn_66,Liu_67,Lynn_68,Cable_69,Lynn_70,Cable_71,Tao_72}.

\subsection{Coexistence of ferroelectric and paraelectric phases}

Similarly to the coexistence of magnetic and paramagnetic phases, many ferroelectrics
exhibit inhomogeneities on mesoscopic scales \cite{Kleemann_57,Bussmann_58,Fu_59}
resulting in the separation of materials onto polarized ferroelectric regions distributed 
randomly within paraelectric mother matrix \cite{Brookeman_60,Gordon_61,Yamada_62}.

The appearance of heterophase fluctuations in the vicinity of ferroelectric-paraelectric 
phase transitions has been studied employing neutron diffuse scattering \cite{Yamada_62}
and M\"{o}ssbauer effect \cite{Bhide_63,Yukalov_64}.

\subsection{Colossal magnetoresistance materials}

In colossal magnetoresistance materials, mesoscopic phase separation plays the key role
for the formation of their specific features \cite{Dagotto_73}. In manganites there happens
the phase separation producing the coexistence of charge ordered metallic phase and 
insulating ferromagnetic phase \cite{Kagan_56,Dagotto_73,Jaime_74,Baio_75,Mathur_76,
Salamon_77,Shimizu_78,Sudheendra_79,Haghiri_80,Mathur_81}. It is this phase coexistence 
and competition that explain the arising colossal magnetoresistance.

\subsection{High-temperature superconductors}

The possibility that in superconductors, due to the existence of Coulomb repulsion,  
there can occur mesoscopic phase separation onto superconducting and normal phases, which
can essentially change the properties of materials, including the superconducting transition 
temperature, was first advanced in Refs. \cite{Shumovsky_82,Yukalov_83}. The model of 
heterophase superconductor \cite{Shumovsky_82,Yukalov_83} was suggested yet before 
high-temperature superconductors were discovered in experiments \cite{Bednorz_84}. Then
this model has been generalized and investigated for superconductors with isotropic gap 
\cite{Yukalov_85,Coleman_86} and anisotropic gap \cite{Yukalov_87,Yukalov_88}. 

The mesoscopic phase separation into a hole-rich superconducting phase and a hole-poor
insulating antiferromagnetic phase has been confirmed in numerous theoretical and 
experimental works \cite{Phillips_89,Benedek_90,Korzhenevskii_91,Muller_92,Sigmund_93,
Nagaev_94,Wubbeler_95,Balagurov_96,Balagurov_97,Pomjakushin_98,Bianconi_99,Kivelson_100}.

The phase separation is dynamical, the arising so-called stripes are self-organized 
networks of charges inside small bubbles of $100-300$ \AA fluctuating at the time scale 
$10^{-12}$ s. The appearance of mesoscopic fluctuations around the phase transition point
in superconductors, leads to the typical anomalous sagging of M\"{o}ssbauer effect factor
\cite{Cherepanov_101}, similarly to the sagging of the M\"{o}ssbauer factor at ferroelectric 
phase transition \cite{Bhide_63,Yukalov_64}. 

It has been suggested \cite{May_102} that the coexistence of superconducting and normal 
phases can also occur in atomic nuclei.

\subsection{Trapped Bose-condensed systems}

In recent years, there has been a high interest to systems with Bose-Einstein condensate 
in traps \cite{Parkins_103,Dalfovo_104,Courteille_105,Andersen_106,Yukalov_107,Bongs_108,
Yukalov_109,Posazhennikova_110,Yukalov_111,Proukakis_112,Yurovsky_113,Yukalov_114,
Yukalov_115,Yukalov_116} and in optical lattices \cite{Morsch_117,Moseley_118,Yukalov_119,
Krutitsky_120,Yukalov_121}. These systems can be in equilibrium (stationary) states as well 
as in nonequilibrium states, when they are subject to external perturbations by modulating 
the trapping potential or atomic interactions \cite{Yukalov_12}. 

By creating strongly nonequilibrium systems, it is possible to form metastable states 
consisting of coexisting condensed and noncondensed phases. Since the Bose-condensed system
is superfluid, the mixture of condensed and noncondensed phases represents coexisting
superfluid and normal phases. Such a metastable state is realized in the form of a
droplet state, where the droplets of Bose condensate float inside incoherent normal gas.
In this case, the coherent droplets of condensate represent mesoscopic fluctuations, as 
far as they are randomly distributed in the trap space, appear and disappear, and have 
the mesoscopic sizes between the mean interatomic distance and the trap lengths. The 
lifetime of a droplet is much longer than the local equilibration time and much shorter
than the existence of the nonequilibrium droplet state of the atomic cloud. The Bose-condensed 
droplet state was observed experimentally \cite{Yukalov_10,Yukalov_11,Bagnato_122,Seeman_123} 
and investigated in numerical calculations 
\cite{Yukalov_10,Yukalov_11,Yukalov_12,Yukalov_124,Yukalov_125,Novikov_126,Yukalov_127}.
A statistical model of the droplet state has been advanced \cite{Yukalov_129}.

\subsection{Fluctuating social groups}

In biological and social systems, there also exist mesoscopic fluctuating groups 
representing group members with different features 
\cite{Crothers_130,Lopez_131,Wallerstein_132}. For example, in a society there are 
cooperators and defectors \cite{Jusup_133} coexisting and competing with each other. 
Competing groups can be formed by the representatives of different social or religious
ideologies \cite{Mises_134}, by the followers of different philosophical or cultural 
trends \cite{Sorokin_135}. In a market, these can be different types of agents, like 
fundamentalists and technical analysts \cite{Edwards_136,Sornette_137}. The competition 
of groups can lead to the development of social instability \cite{Mises_138,Kindleberger_139}
resulting in a kind of social phase transitions \cite{Bouchaud_140,Yukalov_141,Yukalov_142}. 
Similarly to physical systems, group fluctuations often make social systems more stable 
\cite{Yukalov_141,Yukalov_142,Hayek_143}.
 
The typical size of a mesoscopic fluctuating group in a social system can be characterized 
as intermediate in the following sense. The group should consist of many members $N_{mes}$,
but whose number should be essentially smaller than the total number $N$ of the agents in 
the whole social system,
\be
\label{3}
 1 \; \ll \; N_{mes} \; \ll \; N \;   .
\ee
The number of members in a group is not fixed but can fluctuate in time, because of which
the group is labeled as fluctuating.

Since social systems are often represented by physical models, the general methods of 
describing these systems, as well as their mesoscopic fluctuations, are the same as those
used for physical systems.

\section{Spatial and temporal scales}

In statistical systems, it is possible to distinguish several important spatial and temporal 
scales defining the physics of the underlying processes 
\cite{Yukalov_8,Yukalov_9,Yukalov_239}.

The basic spatial scales are the interaction length, or interaction radius, $r_{int}$,
and the mean interparticle distance, $a$. These quantities define the mean particle density
$\rho$ and the characteristic velocity $v$ by the relations
$$
 \rho a^3 \; \approx \; 1 \; , \qquad v \; \sim \; \frac{\hbar}{m r_{int} } \;  ,
$$
where $m$ is particle mass. They also define the mean free path
\be
\label{4}  
 \lbd \; \sim \; \frac{1}{\rho r_{int}^2 } \; \sim \; \frac{a^3}{r_{int}^2}  \;  .
\ee
The size of a mesoscopic fluctuation $l_{mes}$, according to (1), is essentially larger than 
the mean interparticle distance.

The above spatial scales prescribe the characteristic temporal scales. Thus the interaction 
time is
\be
\label{5}
t_{int} \; \sim \; \frac{r_{int}}{v} \; \sim \; \frac{1}{\hbar} \; m r_{int}^2 \; .
\ee
The local equilibration time is
\be
\label{6}
 t_{loc} \; \sim \; \frac{\lbd}{v} \; \sim \; \frac{m a^3}{\hbar r_{int}} \;  ,
\ee
so that its relation with the interaction time becomes
\be
\label{7}
  t_{loc} \; \sim \;  \frac{a^3}{r_{int}^3} \; t_{int} \; .
\ee
The typical lifetime of a mesoscopic fluctuation reads as
\be
\label{8}
t_{mes} \; \sim \; \frac{l_{mes}}{v} \; \sim \; 
\frac{1}{\hbar} \; m r_{int} l_{mes} \; ,
\ee
from where
\be
\label{9}
t_{mes} \; \sim \; \frac{l_{mes}}{r_{int}} \; t_{int} \; \sim \; 
\frac{r_{int}^2 l_{mes}}{a^3} \; t_{loc} \; .
\ee

For condensed matter, $r_{int} \sim a$, hence $\lambda \sim a$, and the mean interparticle 
distance $a \sim 10^{-8}$ cm, while $l_{mes} \gtrsim 10 a$. Then the interaction time is
\be
\label{10}
 t_{int} \; \sim \; \frac{m a^2}{\hbar} \; \sim \; 
10^{-14} - 10^{-13} \; {\rm s} \; ,
\ee
the local equilibration time becomes of the order of the interaction time, 
$t_{loc} \sim t_{int}$, and for the mesoscopic fluctuation time, we have 
\be
\label{11}
  t_{mes} \; \sim \; \frac{l_{mes}}{a}\; t_{loc} \; \gtrsim \; 
10^{-13} - 10^{-12} \; {\rm s} \; .
\ee

In the similar way, it is straightforward to estimate the characteristic quantities for
other systems. For example, for nuclear matter \cite{Yukalov_13,Yukalov_128}, the local 
equilibration and interaction times are of the order of $t_{loc} \sim 10^{-24} - 10^{-23}$ s, 
so that the mesoscopic fluctuation time is $t_{mes} \gtrsim 10^{-23} - 10^{-22}$ s.  
 
Depending on the considered time scale, statistical systems require different types of
description. For a closed system, there are the following relaxation stages. At the 
{\it dynamic stage}
\be
\label{12}
 0 \; < \; t \; < \; t_{int} \qquad (dynamic \; stage) \;  ,
\ee
the system can be strongly nonequilibrium and high-order correlation functions (strictly 
speaking all) are required for its correct description. At the {\it kinetic stage}
\be
\label{13}
 t_{int} \; < \; t \; < \; t_{loc} \qquad (kinetic \; stage) \; ,
\ee
the Bogolubov functional hypothesis \cite{Bogolubov_144} is valid, when all higher-order
correlation functions become functionals of a single-particle distribution function. At 
the {\it heterophase stage}
\be
\label{14}
  t_{loc} \; < \; t \; < \; t_{mes} \qquad (heterophase \; stage) \; ,
\ee
it is necessary to take into account the existence of mesoscopic phase fluctuations. At 
the {\it hydrodynamic stage}
\be
\label{15}
 t_{mes} \; < \; t \; < \; t_{eq} \qquad (hydrodynamic \; stage) \; ,
\ee
the single-particle distribution function can be expressed in terms of time only through 
some of averages, like the average density or energy. After the equilibration time $t_{eq}$,
the system, as a whole, is in equilibrium,
\be
\label{16}
  t_{eq} \; < \; t \; < \; \infty \qquad (equilibrium \; stage) \;  .
\ee
For condensed matter, where $t_{loc} \sim t_{int}$, the kinetic stage is absent. 

Even if the system, as a whole, is equilibrium on large scales, its description depends 
on the intervals of time $\Delta t$ which are considered. On the interval of time
\be
\label{17}
0 \; < \; \Dlt t \; < \; t_{loc} \qquad (nonequilibrium) \;   ,
\ee
the systems is strongly nonequilibrium permanently accomplishing microscopic quantum
and thermal fluctuations. On the mesoscopic time interval
\be
\label{18}
 t_{loc} \; < \; \Dlt t \; < \; t_{mes} \qquad (quasi-equilibrium) \;  ,
\ee
a quasi-equilibrium picture is appropriate considering the arising mesoscopic phase 
fluctuations. For large times, when
\be
\label{19}
 \Dlt t \; \gg \; t_{mes} \qquad (equilibrium) \;   ,
\ee
the overall system can be treated as equilibrium, although taking into account the 
underlying microscopic and mesoscopic fluctuations.

\section{Theory of mesoscopic fluctuations}

There have been a number of phenomenological models attempting to describe the influence
of mesoscopic fluctuations on condensed-matter systems, as can be seen from the references
of review \cite{Yukalov_8}. The ideas of constructing statistical models of systems with 
mesoscopic heterophase fluctuations were advanced in Refs. 
\cite{Yukalov_145,Yukalov_146,Yukalov_147}. In this approach, it is necessary to define the
spaces of states corresponding to different phases, to describe the location of different 
phases by manifold indicator functions \cite{Bourbaki_148,Linnik_149}, as suggested in Refs.
\cite{Yukalov_150,Yukalov_151,Yukalov_152,Yukalov_153}, and to average over phase 
configurations. Here, we follow the general approach developed in Refs. 
\cite{Yukalov_154,Yukalov_155,Yukalov_156,Yukalov_157}. Because of the importance of 
understanding the principal ideas the theory is based on, the main points of the approach 
are explained below. Throughout the paper, the Planck and Boltzmann constants are set to one.

\subsection{Quantum statistical ensemble}

Mesoscopic phase fluctuations exist in both quantum as well as classical systems. The method
of treating them is general for both types of systems. Let us start with the most general 
situation of quantum systems, while returning to classical systems later on. 

A quantum system requires, first, to define the space of accessible microscopic states of 
that system. The microscopic states can be represented by linear combinations over basis 
states $\varphi_n$, with $n$ being a quantum index. Respectively, one says that the system 
is characterized by a Hilbert space
\be
\label{20}
 \cH \; = \; \overline \cL \{ \; \vp_n \; \}  
\ee
that is a closed linear envelope over the given basis. The latter is assumed to be 
orthonormalized. The system state is defined by a statistical operator $\hat{\rho}$ that is
a semi-positive trace-one operator on $\mathcal{H}$. The pair $\{\mathcal{H},\hat{\rho}\}$
is a {\it quantum statistical ensemble}. 
       
Local observables are represented by operators $\hat{A}$ on $\mathcal{H}$ forming the 
algebra of local observables $\mathcal{A}$. The latter is a von Neumann algebra, hence the 
subalgebra of bounded operators on a Hilbert space, which is self-adjoint, closed in the 
weak operator topology, and containing the identity operator. Observable quantities are the
averages
\be
\label{21}
 \lgl \; \hat A \; \rgl \; \equiv \; {\rm Tr}_\cH \hat\rho \; \hat A \; = \;
\sum_n ( \vp_n,\; \hat\rho \hat A \vp_n ) \;  ,
\ee
where the trace operation is over $\mathcal{H}$. The collection of the averages for all 
observable quantities is a {\it statistical state}
\be
\label{22}
 \lgl \; \cA \; \rgl \; =  \; \{ \;  \lgl \; \hat A \; \rgl \; \} \; .
\ee

One usually considers macroscopic statistical systems in the thermodynamic limit, where 
the number of particles $N$ and the system volume $V$ tend to infinity, while the average 
particle density being bounded,
\be
\label{23}
 N \; \ra \; \infty \; , \qquad  
V \; \ra \; \infty \; , \qquad  
\frac{N}{V} \; \ra \; const \; .
\ee
In what follows, when writing $N \ra \infty$, we imply the thermodynamic limit (\ref{23}).

If the considered system can exhibit several thermodynamic phases, enumerated by an index 
$f = 1,2,\ldots$, there occurs the decomposition of observables related to the decomposition 
of the Hilbert space into the direct sum
\be
\label{24}
\cH \; \mapsto \; \bigoplus_f \; \cH_f \qquad ( N \ra \infty)
\ee
of spaces corresponding to different phases. Mathematical details of this decomposition
can be found in Refs. 
\cite{Yukalov_8,Yukalov_9,Ruelle_158,Dixmier_159,Emch_160,Bratelli_161,Palmer_162}.

The structure of the space of states (\ref{24}) means that the system can exist in one of 
the phases, but not in several at once.

\subsection{Methods of phase selection}

In order to study a particular phase $f$, it is necessary to consider not the whole space
$\mathcal{H}$ but only the related subspace $\mathcal{H}_f$. The straightforward way is
to calculate the averages by restricting the trace operation to the required subspace, 
which is called the method of restricted traces \cite{Brout_163,Yukalov_164}. The 
microscopic states of different phases can be distinguished by different symmetry 
properties. Then the selection of a particular phase is named symmetry breaking. Between 
several admissible phases that one is actually realized, which is the most thermodynamically
stable, enjoying, say, the least free energy. This is called spontaneous symmetry breaking. 
Different phases are distinguished by different order parameters or, more generally, by 
different order indices \cite{Coleman_165,Coleman_166,Coleman_167,Coleman_168,Yukalov_169}.  

A convenient way of symmetry breaking, named the {\it method of quasiaverages}, was 
formulated by Bogolubov \cite{Bogolubov_170,Bogolubov_171,Bogolubov_172,Bogolubov_173}.
Suppose the system is characterized by a Hamiltonian $H$ that is invariant with respect to
transformations forming a group. The total Hilbert space of the system is also invariant
with respect to the same group. In order to break the symmetry, it is possible to introduce
the Hamiltonian
\be
\label{25}
H_{f\ep} \; = \; H + \ep \Gm_f
\ee
by adding a term $\Gamma_f$ breaking the symmetry to a subgroup related to the symmetry 
of the required phase, with $\varepsilon$ being a real-valued parameter. The observable 
quantity $A_f$, for a system in phase $f$, corresponding to an operator $\hat{A}$, is 
given by the quasiaverage
\be
\label{26}
 \lim_{N\ra\infty} \; \frac{1}{N} \; A_f \; \equiv \; 
\lim_{\ep\ra 0} \; \lim_{N\ra\infty} \; \frac{1}{N} \; \lgl \; \hat A \; \rgl_{f\ep} \; ,
\ee
where the average on the right-hand side is calculated with Hamiltonian (\ref{25}). Note 
that the limits on the right-hand side are not commutative.  

A convenient modification of this method is provided by the method of thermodynamic 
quasiaverages \cite{Yukalov_153,Yukalov_174}, where, instead of Hamiltonian (\ref{25}),
the Hamiltonian 
\be
\label{27}
H_{f} \; = \; H + \frac{1}{N^\gm} \; \Gm_f \qquad ( 0 < \gm < 1)
\ee
is introduced. The thermodynamic quasi-average is given, instead of two limits in 
(\ref{26}), by the single thermodynamic limit,
\be
\label{28}
A_f \; \equiv \;  \lim_{N\ra\infty} \lgl \; \hat A \; \rgl_{f} \;   ,
\ee
where the average is calculated with Hamiltonian (\ref{27}). 

There exist several other methods of selecting phases, which can be found in Refs.
\cite{Yukalov_8,Brout_163,Sewell_175,Sinai_176}.

\subsection{Weighted Hilbert spaces}

The main idea of all methods for phase selection can be explained by introducing weighted
Hilbert spaces \cite{Yukalov_8,Yukalov_9,Yukalov_177}. Then one defines a probability 
distribution $p(\varphi)$ over the basis $\{\varphi_n\}$, so that
$$
\sum_f p_f(\vp_n) \; = \; 1 \; , \qquad
 0 \; \leq \; p_f(\vp_n) \; \leq \; 1 \;   .
$$
The {\it weighted Hilbert space} is the Hilbert space with the weighted basis,
\be
\label{29}
\cH_f \; \equiv \; \{ \; \cH , \; p_f(\vp) \; \} \;   .
\ee
The observable quantities related to a phase $f$ are given by the averages
\be
\label{30}
 A_f \; \equiv \; {\rm Tr}_{\cH_f} \hat\rho \; \hat A \; \equiv \;
\sum_n  p_f(\vp_n) \; (\vp_n \; , \hat\rho\hat A\vp_n) \; .
\ee

The probability distribution $p_f(\varphi)$ selects the microscopic states typical of 
the required phase $f$ so that to yield the corresponding phase characteristics. If the 
thermodynamic phases are distinguished by the type of symmetry, it is possible to define 
the order operator $\hat\eta$, whose weighted average gives the order parameter
\be
\label{31} 
\eta_f  \; \equiv \; \lgl \; \hat\eta \; \rgl_f \; = \;
{\rm Tr}_{\cH_f} \hat\rho \; \hat\eta
\ee
for each phase $f$. More generally, even when no specific symmetry is ascribed to the 
phases, the latter can be distinguished by the order indices of correlation matrices
\be
\label{32}
\om(\hat C_{nf} )   \; \equiv \; 
\frac{\log||\;\hat C_{nf}\; ||}{\log|\; {\rm Tr} \hat C_{nf}\; | } \;  ,
\ee
where $\hat{C}_{nf}$ is an $n$-th order correlation matrix for the $f$-th phase 
\cite{Coleman_165,Coleman_166,Coleman_167,Coleman_168,Yukalov_169,Yukalov_178}.

\subsection{Spatial phase configuration}

As is clear from the above, the standard approach allows for the consideration of only
pure phases. In order to take into account mesoscopic heterophase fluctuations, it is 
necessary to generalize the approach by including the account of the mesoscopic 
fluctuations.

Assume that we are making a snapshot of the system at a fixed moment of time. The system,
consisting of several phases intermixed in the space, can be characterized by means of 
the manifold indicator functions \cite{Bourbaki_148,Linnik_149} showing the location of
the phases,
\begin{eqnarray}
\label{33}
\xi_f(\br) \; = \; \left\{ \begin{array}{ll}
1 \; , ~ & ~	 \br \; \in \; V_f \\
0 \; , ~ & ~	 \br \; \not\in \; V_f 
\end{array} \right. \; .
\end{eqnarray}
For the simplicity of notation, we denote the spatial part, occupied by a phase $f$, and 
the volume of this part by the same letter $V_f$. The collection of the manifold indicator
functions
\be
\label{34}
\xi \; \equiv \; \{ \; \xi_f(\br) : ~ \br \in V \; , ~ f  = 1,2,\ldots \; \}
\ee
realizes the spatial orthogonal covering due to the orthogonality property
\be
\label{35}
 \xi_f(\br) \; \xi_g(\br)  \; = \; \dlt_{fg} \; \xi_f(\br)
\ee
and the normalization conditions 
\be
\label{36}
 \sum_f \xi_f(\br) \; = \; 1 \; , \qquad
\int_V  \xi_f(\br)  \; d\br \; = \; V_f \; .
\ee
The collection $\xi$ defines the spatial phase configuration.

The phases can occupy different spatial locations and can be of different shapes. This 
variety can be described by introducing for each phase volume the orthogonal subcoverings
with the help of the representation
\be
\label{37}
\xi_f(\br) \; = \; \sum_{i=1}^{n_f} \xi_{fi}( \br-\ba_{fi} ) \;   ,
\ee
separating the phase volume into small cells, where
\begin{eqnarray}
\label{38}
\xi_{fi}(\br) \; = \; \left\{ \begin{array}{ll}
1 \; , ~ & ~	 \br \; \in \; V_{fi} \\
0 \; , ~ & ~	 \br \; \not\in \; V_{fi} 
\end{array} \right. 
\end{eqnarray}
are submanifold indicator functions with the properties of orthogonality
\be
\label{39}
 \xi_{fi}(\br) \; \xi_{fj}(\br)  \; = \; \dlt_{ij} \; \xi_{fi}(\br)  
\ee
and normalization
\be
\label{40}
\int_V  \xi_{fi}(\br)  \; d\br \; = \; V_{fi} \; , \qquad
\sum_{i=1}^{n_f} V_{fi} \; = \; V_f \;   .
\ee
The vectors ${\bf a}_{fi}$ characterize the spatial locations of the cells.

By taking a sufficiently large number $n_f$ of small cells, with the volumes $V_{fi}$,
it is possible to represent any spatial configuration of phases, so that the additive 
Gibbs conditions \cite{Gibbs_179,Ono_180} for the number of particles and volumes be 
valid:
\be
\label{41}
 \sum_f N_f \; = \; N \; , \qquad \sum_f V_f \; = \; V \;  .
\ee

The notion of a separating surface was introduced by Gibbs \cite{Gibbs_179} who emphasized 
that the separating surface is a mathematical abstraction and its exact position can be 
chosen arbitrarily. Its location can be fixed in a unique manner by imposing some additional 
conditions. The separating surface can be located in such a way that the surface density of 
the matter be zero, which implies the additivity of local observables, like in condition 
(\ref{41}). Being an imaginary mathematical abstraction, the Gibbs separating surface should
not be confused with an interface layer having a finite density of matter. A thorough
description of the Gibbs separating surface is given in Ref. \cite{Yukalov_8}.

\subsection{Operators of local observables}  
 
Since the system contains local parts corresponding to different phases, the microscopic 
states of the whole system with coexisting phases pertain to the Hilbert space
\be
\label{42}
 \cF(\cH) \; = \; \bigotimes_f \cH_f \;  ,
\ee
in which the factor spaces are the weighted Hilbert spaces characterizing the related 
phases. The triple, consisting of the total space $\mathcal{F}(\mathcal{H})$, base spaces 
$\mathcal{H}_f$, and the projection map from the total space to a base space is termed 
fibre bundle. The tensor product (\ref{42}) is called the {\it fibred space over a 
Hilbert space} \cite{Husemoller_181,Whitehead_182,Fuks_183}. Note that a similar 
representation exists for Fock spaces \cite{Guichardet_184}. 

The operators of local observables, acting on the fibred space $\mathcal{F}(\mathcal{H})$, 
depend on the phase configuration given by $\xi$ and can be represented by the direct sum
\be
\label{43}
 \hat A(\xi) \; = \; \bigoplus_f \hat A_f(\xi_f) \;  .
\ee
In particular, the operator of the number of particles reads as
\be
\label{44}
  \hat N(\xi) \; = \; \bigoplus_f \hat N_f(\xi_f) \; .
\ee
And the energy Hamiltonian is
\be
\label{45}
 \hat H(\xi) \; = \; \bigoplus_f \hat H_f(\xi_f) \; .
\ee
 
The algebra of operators of local observables, acting on $\mathcal{F}(\mathcal{H})$, 
is denoted as $\mathcal{A}(\xi)$.

\subsection{Heterophase statistical operator}

The statistical operator of a heterophase system $\hat{\rho}(\xi)$ can be defined as 
a minimizer of an information functional, under imposed constraints. First of all, the 
statistical operator has to be normalized,
\be
\label{46}
{\rm Tr} \int \hat\rho(\xi) \; \cD \xi \; = \; 1 \;  ,
\ee
where the trace is over the fibred space (\ref{42}) and the integral over $\xi$ implies
a functional integration over the set (\ref{34}) of the manifold indicator functions. This
integration symbolizes the averaging over phase configurations. The averaging is required
due to the mesoscopic nature of the phase fluctuations, whose lifetime is much shorter
than the observation time and whose locations are randomly distributed over the sample.  

The second constraint is the definition of the average energy
\be
\label{47}
{\rm Tr} \int \hat\rho(\xi) \; \hat H(\xi) \; \cD \xi \; = \;  E \;   .
\ee
The given total number of particles in the system implies the condition
\be 
\label{48}
{\rm Tr} \int \hat\rho(\xi) \; \hat N(\xi) \; \cD \xi \; = \;  N \;   .
\ee
Similarly, one can impose other constraints uniquely defining a representative ensemble
\cite{Yukalov_8,Yukalov_9,Yukalov_185,Yukalov_186,Yukalov_193}.   

The general information functional has the Kullback-Leibler \cite{Kullback_187,Kullback_188}  
form
$$
I[\; \hat\rho\; ] \; = \;  {\rm Tr} \int \hat\rho(\xi) \; 
\ln \; \frac{\hat\rho(\xi)}{\hat\rho_0(\xi) } \; \cD\xi +
\al \left[ \; {\rm Tr} \int \hat\rho(\xi) \; \cD\xi - 1 \; \right] \; +
$$
\be
\label{49}
+ \; 
\bt \left[ \; {\rm Tr} \int \hat\rho(\xi) \; \hat H(\xi) \cD\xi - E \; \right] +
\gm \left[ \; {\rm Tr} \int \hat\rho(\xi) \; \hat N(\xi) \cD\xi - N \; \right] \; ,
\ee
in which $\alpha$, $\beta$, and $\gamma$ are Lagrange multipliers and $\hat{\rho}_0(\xi)$ 
is a prior statistical operator imposing restrictions on the distribution of phase 
configurations, if these are known. In the case that no prior information on the random
distribution of configurations is available, the prior statistical operator is assumed 
to be constant.   

Minimizing the information functional yields the {\it heterophase statistical operator}
\be
\label{50}
\hat\rho(\xi) \; = \; 
\frac{\hat\rho_0(\xi)\exp\{-\bt H(\xi)\} }{{\rm Tr}\hat\rho_0(\xi)\exp\{-\bt H(\xi)\} } \; ,
\ee
with the chemical potential $\mu = -\gamma T$, $\beta = 1/T$, temperature $T$, and the 
grand Hamiltonian
\be
\label{51}
H(\xi) \; \equiv \; \hat H(\xi) - \mu \; \hat N(\xi) \;   .
\ee
When there is no information on the trial distribution of the phases, the statistical 
operator becomes
\be
\label{52}
\hat\rho(\xi) \; = \; \frac{1}{Z} \; \exp\{ -\bt H(\xi) \; \} \; ,
\ee
with the partition function
\be
\label{53}
Z \; = \; {\rm Tr} \int \exp\{\; - \bt H(\xi) \; \} \; \cD\xi \; .
\ee

In the second-quantization representation, the energy Hamiltonian of the phase $f$, taking 
into account the identity
$$
 \int_{V_f} d\br \; = \; \int_V \xi_f(\br) \; d\br \;  ,
$$
reads as
$$
\hat H_f(\xi_f) \; = \; \int \xi_f(\br) \; \psi_f^\dgr(\br) \; \left[ \; - \; 
\frac{\nabla^2}{2m} + U(\br) \; \right] \; \psi_f(\br) \; d\br \; +
$$
\be
\label{54}
+ \;
 \frac{1}{2} \int \xi_f(\br) \; \xi_f(\br') \; \psi_f^\dgr(\br) \; \psi_f^\dgr(\br') \; 
\Phi(\br-\br') \; \psi_f(\br') \; \psi_f(\br) \; d\br d\br'
\ee
and the number-of-particle operator of the phase $f$ is
\be
\label{55}
\hat N_f(\xi_f) \; = \; \int \xi_f(\br) \; \psi_f^\dgr(\br) \; \psi_f(\br) \; d\br \;  .
\ee
Here $\psi_f({\bf r})$ is the representation of a field operator on the Hilbert space 
$\mathcal{H}_f$; $U({\bf r})$ is an external potential, and $\Phi({\bf r}$ is an 
interaction potential. Here and in what follows, where the integration over the space 
is not shown, it is implied that the integration is over the whole system volume $V$. 
The internal degrees of freedom, such as spin, isospin, or like that, can be taken into 
account by representing the field operators as columns whose rows are labeled by the 
internal degrees of freedom.

\subsection{Averaging over phase configurations}

Technically, the averaging over phase configurations is accomplished by defining the 
integration over the manifold indicator functions. For this purpose, we introduce the 
variable
\be
\label{56}
x_f \; = \; \frac{1}{V} \int \xi_f(\br) \; d\br \; = \; \frac{V_f}{V}
\ee
that varies in the interval $[0,1]$, satisfying the normalization condition
\be
\label{57}
  \sum_f x_f \; = \; 1 \; , \qquad 0 \; \leq \; x_f \; \leq 1 \; .
\ee
As is explained in Sec. 4.4, the overall system volume can be partitioned into small 
cells whose locations are described by the submanifold indicator functions (\ref{38}). 
The functional integration consists of two steps. First, by changing the cell locations 
it is possible to create various phase configurations, provided the normalization 
(\ref{57}) is valid and the cells are asymptotically small, such that
\be
\label{58}
 n_f \; \ra \; \infty \; , \qquad V_{fi} \; \ra \; 0 \; , \qquad x_f \; = \; const \;  .
\ee
At the second step, the phase concentrations are to be varied, which means the variation
of $x_f$ in the range $[0,1]$. 

Thus the differential measure of the functional integration is
\be
\label{59}
 \cD \xi \; = \; \dlt\left( \sum_f x_f - 1 \right) \; 
\prod_f d x_f \; \prod_f \cD \xi_f \;  ,
\ee
with 
\be
\label{60}
\cD \xi_f \; = \; \prod_{i=1}^{n_f} \frac{d\ba_{fi}}{V} 
\qquad 
( n_f \; \ra \; \infty ) \;   .
\ee
The functional integration over the manifold indicator functions yields the following 
results \cite{Yukalov_8,Yukalov_154,Yukalov_155,Yukalov_156,Yukalov_157}.

\vskip 2mm

{\bf Lemma}. Let us consider the class $\mathcal{C}_f$ of the functionals having the structure
\be
\label{61}
 C_f(\xi_f) \; = \; \sum_n \int C_f(\br_1,\br_2,\ldots,\br_n) \;
\prod_{j=1}^n \xi_f(\br_j) \; d\br_j \;  .
\ee
The averaging over phase configurations, under fixed concentrations $x_f$, gives
\be
\label{62}
\int  C_f(\xi_f) \; \cD \xi_f \; = \; C_f(x_f) \;  ,
\ee
where
\be
\label{63}
C_f(x_f) \; = \; \sum_n x_f^n \; 
\int C_f(\br_1,\br_2,\ldots,\br_n) \; d\br_1 d\br_2 \ldots d\br_n \;  .
\ee
 
\vskip 2mm
{\bf Theorem 1}. Let the representation of the Hamiltonian on the space $\mathcal{H}_f$
pertain to the class $\mathcal{C}_f$, having the form
\be
\label{64}
 H_f(\xi_f) \; = \; \sum_n H_f(\br_1,\br_2,\ldots,\br_n) \; 
\prod_{j=1}^n \xi_f(\br_j) \; d\br_j \; .
\ee
Then it follows that
\be
\label{65}
 \int \exp\{- \bt H_f(\xi_f) \} \; \cD \xi_f \; = \;
\exp \{- \bt H_f(x_f) \} \; ,
\ee
where
\be
\label{66}
H_f(x_f) \; = \; \sum_n x_f^n \; 
\int H_f(\br_1,\br_2,\ldots,\br_n) \; d\br_1 d\br_2 \ldots d\br_n \; .
\ee

\vskip 2mm
{\bf Theorem 2}. Let the thermodynamic potential be
\be
\label{67}
\Om \; = \; - T\ln Z \; = \; - 
T \ln {\rm Tr} \int \exp\{ \; - \bt H(\xi) \; \} \; \cD \xi \;  ,
\ee
with the Hamiltonian 
\be
\label{68}
  H(\xi) \; = \; \bigoplus_f H_f(\xi_f) 
\ee
acting on the fibred space (\ref{42}), with $H_f(\xi_f)$ being of the class $\mathcal{C}_f$, 
as defined in (\ref{64}). Then the averaging over phase configurations yields
\be
\label{69}
\Om \; = \; {\rm abs} \; \min_{\{ w_f\} } \Om(\{ w_f \} ) \;   ,
\ee
under the normalization condition
\be
\label{70}
 \sum_f w_f \; = \; 1 \; , \qquad 0 \; \leq \; w_f \; \leq \; 1 \;  .
\ee
Here
\be
\label{71}
  \Om(\{ w_f \} ) \; = \;  -T \ln {\rm Tr} \; \exp\{ - \bt H_{eff} \} \;  ,
\ee
with the effective Hamiltonian 
\be
\label{72}
 H_{eff} \; = \; \bigoplus_f H_f ( w_f) \;  ,
\ee
where
\be
\label{73}
H_f(w_f) \; = \;  \sum_n w_f^n \; 
\int H_f(\br_1,\br_2,\ldots,\br_n) \; d\br_1 d\br_2 \ldots d\br_n \; .
\ee
In other words,
\be
\label{74}
 \Om \; = \; \sum_f \Om_f(w_f) \; , \qquad
\Om_f(w_f) \; = \; -T \ln {\rm Tr}_{\cH_f} \; \exp\{ - \bt H_f(w_f) \} \;  .
\ee
The quantity $w_f = V_f/V$, minimizing the thermodynamic potential, plays the role of 
{\it phase probability}.   

\vskip 2mm
{\bf Theorem 3}. The observable quantity for an operator (\ref{43}), with the terms of 
the class $\mathcal{C}_f$,
\be
\label{75}
 \hat A_f(\xi_f) \; = \; 
\sum_n \int A_f(\br_1,\br_2,\ldots,\br_n) \; \prod_{j=1}^n \xi_f(\br_j) \; d\br_j \; ,
\ee
given by the average
\be
\label{76}
 \lgl \; \hat A \; \rgl \; = \; 
{\rm Tr} \int \hat\rho(\xi) \; \hat A(\xi) \; \cD \xi \;  ,
\ee
with the statistical operator 
\be
\label{77}
 \hat\rho(\xi) \; = \; \frac{1}{Z} \; \exp \{ - \bt H(\xi) \} \;  ,
\ee
after averaging over phase configurations, reduces to 
\be
\label{78}
 \lgl \; \hat A \; \rgl \; = \; \sum_f  \lgl \; \hat A_f(w_f) \; \rgl \;  ,
\ee
where
\be
\label{79}
\lgl \; \hat A_f(w_f) \; \rgl \; \equiv \; 
{\rm Tr}_{\cH_f} \hat\rho_f(w_f) \; \hat A_f(w_f) \;  ,
\ee
with the statistical operator
\be
\label{80}
 \hat\rho_f(w_f) \; = \; \frac{1}{Z_f} \; \exp\{ -\bt H_f(w_f) \} \;   ,
\ee
the partition function
\be
\label{81}
 Z_f \; = \;  {\rm Tr}_{\cH_f}  \exp\{ -\bt H_f(w_f) \} \; ,
\ee
and the renormalized terms
\be
\label{82}
 \hat A_f(w_f) \; = \; \sum_n w_f^n \; 
\int A_f(\br_1,\br_2, \ldots,\br_n) \; d\br_1 d\br_2 \ldots d\br_n \;   .
\ee

\vskip 2mm
For example, the renormalized expression for Hamiltonian (\ref{54}) is
$$
H_f(w_f) \; = \; w_f \int \psi_f^\dgr(\br) \; \left[ \; - \; 
\frac{\nabla^2}{2m} + U(\br) \; \right ] \; \psi_f(\br) \; d\br \; +
$$
\be
\label{83}
+ \; \frac{1}{2} \; w_f^2 \int \psi_f^\dgr(\br) \; \psi_f^\dgr(\br') \; 
\Phi(\br-\br') \; \psi_f(\br') \; \psi_f(\br) \; d\br d\br' \; ,
\ee
while the renormalized number-of-particle operator (\ref{55}) becomes
\be
\label{84}
 \hat N_f(w_f) \; = \; 
w_f \int \psi_f^\dgr(\br) \; \psi_f(\br) \; d\br \;  .
\ee
 
It is easy to notice that if mesoscopic fluctuations are absent, and just a single phase 
can exist, we return to the standard formulation for describing statistical systems.

\subsection{Surface thermodynamic potential}

As has been emphasized above, the Gibbs separating surface is an imaginary dividing
boundary between the phases, and no surface operators or other microscopic quantities 
exist at the microscopic level. But surface characteristics appear at the macroscopic 
level, when studying thermodynamic properties. Thus the surface thermodynamic potential 
is defined \cite{Gibbs_179,Kjelstrup_189} as the difference 
\be
\label{85}
\Om_{sur} \; \equiv \; \Om - \Om_G 
\ee
between the actual thermodynamic potential of a heterophase system in the volume $V$ and 
the thermodynamic potential of the Gibbs mixture, which is the sum of the thermodynamic 
potentials of pure phases occupying the related volumes $V_f$,
\be
\label{86}
 \Om_G \; = \; \sum_f w_f \; \Om_f(1) \; .
\ee
Hence the surface thermodynamic potential is
\be
\label{87}
 \Om_{sur} \; = \; \Om - \sum_f w_f\; \Om_f(1) \;  .
\ee
By definition (\ref{69}), we have
\be
\label{88}
 \Om \; = \; 
{\rm abs} \; \min_{\{ w_f \} } \Om(\{ w_f \} ) \; \leq \; \min_f \Om_f(1) \; .
\ee
Moreover,
\be
\label{89}
 \min_f \Om_f(1) \; \leq \; \sum_f w_f \; \Om_f(1) \;  ,
\ee
because of which
\be
\label{90}
   \Om \; \leq \; \sum_f w_f \; \Om_f(1) \; .
\ee
Then from (\ref{87}) it follows that the heterophase system is stable when
\be 
\label{91}
\Om_{sur} \; \leq \; 0 \;  .
\ee

In the similar way, it is straightforward to define the surface quantities corresponding 
to the averages for the operators of observables, as the difference between the actual 
average for a heterophase system and the sum of the averages for the Gibbs mixture 
components,
\be
\label{92}
\lgl \; \hat A \; \rgl_{sur} \; \equiv \;  
\lgl \; \hat A \; \rgl - \sum_f w_f \lgl \; \hat A_f(1) \; \rgl \; ,
\ee
which gives
\be
\label{93}
\lgl \; \hat A \; \rgl_{sur} \; = \;  \sum_f \left[ \;
\lgl \; \hat A_f(w_f) \; \rgl - w_f \lgl \; \hat A_f(1) \; \rgl \; \right] \;  .
\ee
The set of surface observables (\ref{93}) constitutes the surface state. Notice that the 
surface number of particles is zero, $\langle \hat{N}(\xi) \rangle_{sur} = 0$, in agreement 
with the definition of the Gibbs separating surface. However the surface energy
$\langle \hat{H}(\xi) \rangle_{sur}$ is not zero, since the renormalized Hamiltonian is 
not linear in the phase probabilities $w_f$.

\subsection{Thermodynamics of heterophase systems}

All thermodynamic characteristics can be calculated as soon as we find a thermodynamic 
potential. For concreteness, we have considered above the grand thermodynamic potential.
Other thermodynamic potentials can be obtained either by employing Legendre transformations
\cite{Yukalov_190} or by directly calculating the required potentials following the same 
scheme as above. For instance, for the free energy of a heterophase system, we find
\be
\label{94}   
F \; = \; \sum_f F_f(w_f) \;   ,
\ee
with the terms
\be
\label{95}
 F_f(w_f) \; = \; - T \ln {\rm Tr}_{\cH_f} \exp\{ -\bt \hat H_f(w_f) \} \;  .
\ee
The surface free energy reads as
\be
\label{96}
 F_{sur} \; = \; F - \sum_f w_f \; F_f(1) \;  .
\ee

The principal difference from the description of pure phases is the existence of the 
additional thermodynamic quantities called {\it geometric phase probabilities}
\be
\label{97}
 w_f \; = \; \frac{V_f}{V} \qquad ( f = 1,2,\ldots)  
\ee
that are defined as minimizers of a thermodynamic potential. The phase probabilities
appear due to the occurrence of mesoscopic phase fluctuations. It may happen that these
fluctuations exist not for all thermodynamic parameters but arise under the variation
of one or several parameters. For instance, it may happen that below some temperature
$T_{nuc}$, the system consists of a pure phase, say phase $f$, so that 
\be
\label{98}
 w_f \; = \;  1 \qquad ( T < T_{nuc} ) \;   ,
\ee
while above this temperature the system becomes heterophase, when
\be
\label{99}
w_f \;  < \; 1 \qquad ( T > T_{nuc} ) \;    .
\ee
The corresponding temperature is called the {\it nucleation temperature}.

Except the phase probabilities, it is possible to define the phase fractions
\be
\label{100}
n_f \; \equiv \;  \frac{N_f}{N} \qquad ( f = 1,2,\ldots) \; ,
\ee
which are related with the phase probabilities through the phase density
\be
\label{101}
\rho_f \; \equiv \;  \frac{N_f}{V_f} \; = \;  \frac{n_f}{w_f}\; \rho
\ee
and the average particle density 
\be
\label{102}
 \rho \; \equiv \;  \frac{N}{V} \; = \;  \sum_f w_f\; \rho_f \;  .
\ee
If the phases do not differ by their densities, then the phase probabilities and phase
fractions coincide with each other,
\be
\label{103}
 w_f \; = \;  n_f \qquad ( \rho_f = \rho ) \;   .
\ee

\section{Classical heterophase systems}
 
In the previous section, the general approach for taking account of mesoscopic heterophase 
fluctuations in quantum systems is presented. Of course, these fluctuations are not a 
privilege of only quantum systems, but do exist in classical systems as well. The overall
approach can be straightforwardly reformulated for classical systems, as is sketched below.

\subsection{Classical statistical ensemble}

A classical statistical system of many particles $N$ is characterized by the position and
momentum variables 
\be
\label{104}
q \; = \; \{\bq_1, \bq_2, \ldots,\bq_N\} \; , \qquad
p \; = \; \{\bp_1, \bp_2, \ldots,\bp_N\} \;    .
\ee
In total, for a $d$-dimensional space, there are $2Nd$ variables, whose collection 
$\{q,p\}$ forms the {\it space of microstates}. This space is made measurable introducing
the differential measure
\be
\label{105}
d\mu(q,p) \; = \; \frac{dq\; dp}{N!(2\pi\hbar)^{3N} } \;   .
\ee
Then the measurable space of microstates constitutes the {\it phase space}
\be
\label{106}
{\cal M} \; = \; \{ \; q,p,\mu(q,p)\; \}  .
\ee
Defining a probability distribution $\rho(q,p)$ makes the pair $\{\mathcal{M}, \rho(q,p)\}$
a {\it classical statistical ensemble}.    

The observables are given by the functionals $A(q,p)$ whose averages 
\be
\label{107}
\lgl \; A \; \rgl \; = \; \int A(q,p) \; \rho(q,p) \; d\mu(q,p)
\ee
over the phase space yield observable quantities. The collection of the latter defines
a {\it classical statistical state}. 

Generally, the total phase space can characterize different thermodynamic phases, being the 
direct sum of subspaces corresponding to particular pure accessible phases,
\be
\label{108}
{\cal M} \; = \; \bigoplus_f {\cal M}_f \; .
\ee
Depending on the given thermodynamic parameters, that pure phase is actually realized that 
is the most thermodynamically stable.

\subsection{Selection of thermodynamic phases}

In order to select a required thermodynamic phase, it is necessary to restrict the phase 
space, thus defining restricted averages \cite{Yukalov_8}, similarly to the use of 
restricted traces for quantum systems. The general way is to employ weighted phase spaces
\cite{Yukalov_191}, which is analogous to employing weighted Hilbert spaces in quantum 
theory. For this purpose, one has to define a weighting function $\nu_f(q,p)$, 
such that
\be
\label{109}
\sum_f \nu_f(q,p) \; = \; 1 \; , \qquad 0 \; \leq \; \nu_f(q,p) \; \leq \; 1 \;  ,
\ee
whose role is to select the part of the phase space typical of the needed thermodynamic
phase. The phase space complimented by the weighting function forms the {\it weighted 
phase space}
\be
\label{110}
 {\cal M}_f \; = \; \{ \; {\cal M}, \; \nu_f(q,p) \; \} \;  .
\ee
Introducing the weighted differential measure
\be
\label{111}
 d\mu_f(q,p) \; = \; \nu_f(q,p) \; d\mu(q,p)  
\ee
allows one to characterize the chosen phase by the averages related to the observable
quantities
\be
\label{112}
A_f \; = \; \int A(q,p) \; \rho(q,p) \; d\mu_f(q,p) \;  .
\ee

\subsection{Classical heterophase probability distribution}

The following consideration of classical systems with mesoscopic phase fluctuations is 
completely similar to the quantum case. A heterophase system is characterized by the
fibred phase space
\be
\label{113}
 {\cal F}({\cal M}) \; = \; \bigotimes_f {\cal M}_f \;  .
\ee
Introducing the manifold indicator functions, showing the localization of different phases,
one obtains the local observables $A_f(q,p,\xi_f)$. The observable quantities are given by
the averages of the functionals
\be
\label{114}
A(q,p,\xi) \; = \; \bigoplus_f A_f(q,p,\xi_f)
\ee
over the weighted phase spaces and the averaging over phase configurations,
\be
\label{115}
 \lgl \; A \; \rgl \; = \; 
\int A(q,p,\xi) \; \rho(q,p,\xi) \; d\mu(q,p) \; \cD\xi \;  ,
\ee
where
\be
\label{116}
d\mu(q,p) \; = \; \bigotimes_f d\mu_f(q,p) \;   .
\ee
   
The system Hamiltonian has the form
\be
\label{117}
 H(q,p,\xi) \; = \; \bigoplus_f H_f(q,p,\xi_f) \; ,
\ee
with the terms
\be
\label{118}
 H_f(q,p,\xi_f) \; = \; \sum_{i=1}^N \xi_f(\br_i) \; \left[ \; 
\frac{\bp^2}{2m} + U(\br_i) \; \right] + \frac{1}{2}
\sum_{i\neq j}^N \xi_f(\br_i) \; \xi_f(\br_j) \; \Phi(\br_i - \br_j) \; .
\ee

The probability distribution has to satisfy the normalization condition
\be
\label{119}
\int \rho(q,p,\xi) \; d\mu(q,p) \; \cD\xi \; = \; 1
\ee
and the definition of the average energy
\be
\label{120}
 \int \rho(q,p,\xi) \; H(q,p,\xi) \; d\mu(q,p) \; \cD\xi \; = \; E \; .
\ee

The information functional reads as
$$
I[\; \rho \; ] \; = \; \int \rho(q,p,\xi) \; \ln \;
\frac{\rho(q,p,\xi)}{\rho_0(q,p,\xi)} \; d\mu(q,p) \; \cD\xi \; + \;
$$
\be
\label{121}
+ \; 
\al\; \left[\; \int \rho(q,p,\xi) \; d\mu(q,p) \; \cD\xi - 1 \; \right] +
\bt\; \left[\; 
\int \rho(q,p,\xi) \; H(q,p,\xi) \; d\mu(q,p) \; \cD\xi - E \; \right] \; .
\ee
If there is no prior information on the distribution of heterophase fluctuations, the 
trial distribution $\rho_0(q,p,\xi)$ is a constant. Then the minimization of the 
information functional results in the probability distribution
\be
\label{122}
 \rho(q,p,\xi) \; = \; \frac{1}{Z} \; \exp\{ -\bt H(q,p,\xi) \} \;  ,
\ee
with the partition function
\be
\label{123}
Z \; = \; \int\exp\{ - \bt H(q,p,\xi) \} \; d\mu(q,p) \; \cD\xi \; .
\ee

\subsection{Results of averaging over phase configurations}

The averaging over phase configurations, realized as the functional integration over 
the manifold indicator functions, is accomplished in the same way as is described in 
the previous sections above. As a result, we come to the following picture.

The effective Hamiltonian takes the form
\be
\label{124}
H_{eff}(q,p) \; = \; \bigoplus_f H_f(q,p,w_f) \;  ,
\ee
with the terms
\be
\label{125}
H_f(q,p,w_f) \; = \; w_f \sum_{i=1}^N \left[ \;
\frac{\bp_i^2}{2m} + U(\br_i) \; \right] + \frac{1}{2} \; w_f^2
\sum_{i\neq j}^N \Phi(\br_i - \br_j) \; .
\ee

The probability distribution is
\be
\label{126}
\rho(q,p) \; = \; \bigotimes_f \rho_f(q,p,w_f) \; ,
\ee
with the factors
\be
\label{127}
\rho_f(q,p,w_f) \; = \; \frac{1}{Z_f} \; \exp\{ -\bt H_f(q,p,w_f) \}    
\ee
and the partition functions
\be
\label{128}
 Z_f \; = \; \int \exp\{ - \bt H_f(q,p,w_f) \} \; d\mu_f(q,p) \; .
\ee
If $\nu_f(q,p) = \nu_f(q)$ does not depend on momenta, then, after the integration over 
momenta, one has
\be
\label{129}
Z_f \; = \; \left( \frac{mT}{2\pi w_f}\right)^{3N/2} \frac{1}{N!}
\int \exp\left\{ \; -\; \frac{w_f^2}{2T} \sum_{i\neq j}^N \Phi(\br_i - \br_j) \; - \; 
\frac{w_f}{T} \sum_{i=1}^N U(\br_i) \; \right\} \; \nu_f(q) \; d q \;  .
\ee

The observable quantities are defined by the averages
\be
\label{130}
\lgl \;  A \; \rgl \; = \; \sum_f \lgl \; A_f \; \rgl \; ,
\ee
where
\be
\label{131}
 \lgl \;  A_f \; \rgl \; = \; 
\int \rho_f(q,p,w_f) \; A_f(q,p,w_f) \; d\mu_f(q,p) \;  .
\ee

\subsection{Quasiaverages and weighted spaces}

The choice of a weighted phase space, required for describing particular 
thermodynamic phases, can be realized in different ways \cite{Yukalov_8,Yukalov_9} 
allowing for the definition of restricted phase spaces for classical systems 
\cite{Frenkel_5} or restricted traces for  quantum systems \cite{Brout_163}. 
A convenient method is the method of quasiaverages that was mentioned by 
Kirkwood \cite{Kirkwood_192} and developed by Bogolubov 
\cite{Bogolubov_171,Bogolubov_172,Bogolubov_173}. Below we show that the 
method of quasiaverages is a particular case of the method of weighted spaces. 

The average (\ref{131}) can be written in the form
\be
\label{132}
 \lgl \; \hat A_f \; \rgl \; = \; 
\frac{\int\exp\{-\bt H_f(q,p)\} A_f(q,p) \; d\mu_f(q,p)}
     {\int\exp\{-\bt H_f(q,p)\} \; d\mu_f(q,p)} \;  .
\ee

In the method of quasiaverages, one introduces a Hamiltonian
\be
\label{133}
H_{f\ep}(q,p) \; \equiv \; H_f(q,p) + \ep \; \Gm_f(q,p)
\ee
by adding a term breaking the symmetry of the Hamiltonian $H_f$, and reducing it to 
a subgroup of symmetry characterizing the phase $f$. The quasiaverages are defined 
by the relation
\be
\label{134}
 \lim_{N\ra\infty} \; \frac{1}{N} \; \lgl \; A_f \; \rgl \; = \;
\lim_{\ep\ra 0} \; \lim_{N\ra\infty} \;
\frac{\int\exp\{-\bt H_{f\ep}(q,p)\} A_f(q,p) \; d\mu_f(q,p)}
     {N \int\exp\{-\bt H_{f\ep}(q,p)\} \; d\mu_f(q,p)} \; ,
\ee
where the limit $\varepsilon \ra 0$ is taken after the thermodynamic limit. Comparing 
(\ref{132}) and (\ref{134}), we see that they coincide provided that in (\ref{132}) one 
defines the weight 
\be
\label{135}
 \nu_f(q,p) \; = \; \exp\{ - \bt \ep \Gm_f(q,p) \} \;  ,
\ee
keeping in mind that $\varepsilon \ra 0$ after the thermodynamic limit.

\section{Models of heterophase systems}

The theory of systems with mesoscopic phase fluctuations is illustrated below by several 
examples. The derivation of effective Hamiltonians, obtained by averaging over heterophase 
configurations, has been explained above.

\subsection{Magnetic materials with paramagnetic fluctuations}

Let us consider a ferromagnetic material on a lattice with a fixed spatial structure, 
where there can arise paramagnetic fluctuations. The effective Hamiltonian has the 
form
\be
\label{136}
H \; = \; H_1 \bigoplus H_2 \; ,
\ee
in which 
\be
\label{137}
H_f \; = \; \frac{1}{2} \; w_f^2 \; U N - 
w_f^2 \sum_{i\neq j} J_{ij} \; \bS_{if} \cdot \bS_{jf} \; .
\ee
Here $U$ is the average strength of direct particle interactions, while $J_{ij}>0$ 
is an exchange interaction potential. Note that the term containing the direct 
interactions $U$ cannot be omitted, since it depends of the phase probabilities 
$w_f$. This term is principal for the existence of paramagnetic fluctuations.    

The order parameter is
\be
\label{138}
 s_f \; \equiv \; \frac{1}{N S}   \sum_{j=1}^N \; \lgl \; S_{jf}^z \; \rgl
\ee
under the condition 
\be
\label{139}
 s_1 \; > \; 0 \; , \qquad  s_2 \; = \; 0 \; , 
\ee
classifying the first phase as ferromagnetic, with the second phase corresponding 
to paramagnetic fluctuations. The lattice is assumed to be fixed and being sufficiently 
deep, so that all spins are well localized in their lattice sites.  

The model of such a ferromagnet with mesoscopic phase fluctuations, and its 
properties, have been studied, in mean-field approximation, in Refs.
\cite{Yukalov_49,Shumovsky_194,Shumovsky_195,Shumovsky_196,Shumovsky_197,Shumovsky_198}. 
Depending on the parameter 
\be
\label{140}
u \; \equiv \; \frac{U}{J} \qquad 
\left( J \equiv \frac{1}{N} \sum_{i\neq j}^N J_{ij} \right) \; ,
\ee
there can occur the following cases. 

For $u \leq 0$, paramagnetic fluctuations cannot appear. The sample is purely ferromagnetic  
below the critical temperature $T_c = 0.5$, where by the phase transition of second order 
the sample turns into paramagnetic phase.

For $0 < u < 0.5$, the system is purely ferromagnetic below the temperature $T_0$, where 
it transforms into paramagnetic phase by the phase transition of first order. The transition
temperature is in the interval $0.09 < T_0 < 0.18$. 

For $0.5 \leq u < 1.5$, in the region below $T_0$, the system is in the mixed phase of a 
ferromagnet with paramagnetic fluctuations. The mixed phase transforms into paramagnetic 
phase by the first-order phase transition at temperature $T_0$ in the range 
$0.125 < T_0 < 0.182$.

For $u = 1.5$, the heterophase system is ferromagnetic with paramagnetic fluctuations 
below the tricritical point $T_c^* = 0.182$. The tricritical point lays on the boundary 
between first and second order phase transitions \cite{Lawrie_199}.  

For $u > 1.5$, the sample is heterophase, being a ferromagnet with paramagnetic fluctuations
below the critical temperature $T_c = 0.125$, where it becomes nonmagnetic through the 
second-order phase transition. 

In the random-phase approximation, the transition between the ferromagnet with paramagnetic
fluctuations and a paramagnet, for realistic parameters, occurs through first-order phase
transition \cite{Yukalov_200}. 

Several other models of magnets with mesoscopic fluctuations have been considered, such as 
antiferromagnets with paramagnetic fluctuations \cite{Boky_201}, heterophase Hubbard model
\cite{Boky_222}, heterophase Vonsovsky-Zeener model \cite{Yukalov_49}, heterophase 
ferromagnets with phonon excitations \cite{Yukalov_202}, heterophase generalization of the 
Nagle \cite{Nagle_203,Kislinsky_204} model with short- and long-range interactions 
\cite{Yukalov_205}, heterophase spin-glass model with paramagnetic fluctuations 
\cite{Yukalov_205}, and a ferromagnet with magnetic reorientations and paramagnetic 
fluctuations \cite{Bakasov_206,Bakasov_207,Yukalov_208}. Two-dimensional heterophase 
Ising-type model, describing a ferromagnet with paramagnetic fluctuations turns out to be 
metastable \cite{Kislinsky_209}. These models were considered at zero magnetic field. 
Switching on a nonzero external magnetic field leads to the suppression of mesoscopic 
paramagnetic fluctuations in magnets \cite{Yukalov_210}.

\subsection{Ferroelectrics with paraelectric fluctuations}

There are numerous examples of ferroelectrics exhibiting paraelectric fluctuations
\cite{Kleemann_57,Bussmann_58,Fu_59,Brookeman_60,Gordon_61,Yamada_62,Cook_211,Rigamonti_212,
Gordon_213,Gordon_214}. We consider ferroelectrics with the phase transitions of the 
order-disorder type \cite{Blinc_215}. Atoms or molecules, composing a ferroelectric, are
located on a lattice where each site is formed by a double well. Electric dipoles appear 
due to atomic displacement inside double wells. Taking into account paraelectric fluctuations
leads \cite{Yukalov_216} to the effective Hamiltonian $H = H_1 \bigoplus H_2$  with the 
partial Hamiltonians
$$
H_f \; = \; w_f E_0 - w_f \sum_j \left( \Om_j S_{jf}^x + D_j S_{jf}^z \right) \;
+
$$
\be
\label{141}
+ \;
w_f^2 \sum_{i\neq j} \left( \frac{1}{2} \; U_{ij} + B_{ij} S_{if}^x S_{jf}^x - 
J_{ij} S_{if}^z S_{jf}^z \right) \; .
\ee
Here, the first term, linear in $w_f$, can be omitted, since in the total Hamiltonian it 
gives a constant $w_1 E_0 + w_2 E_0 = E_0$. The second term describes the tunneling between 
the wells of a double-well potential at the site $j$, with the tunneling frequency $\Om_j$. 
The third term is caused by strain energy due to external forces or to the action of 
electric field. The quantity $U_{ij}$ is a matrix element of direct particle interactions, 
while $B_{ij}$ and $J_{ij}$ are the matrix elements of exchange interactions, transverse 
and longitudinal. Details on the parameters of the model are given in Refs. 
\cite{Yukalov_217,Yukalov_218}. Similar Hamiltonians describe atoms in optical lattices 
\cite{Yukalov_219,Yukalov_220} and some macromolecular systems \cite{Yukalov_221}. 

The order parameters characterize the mean atomic imbalance
\be
\label{142}
s_f \; \equiv \; \frac{2}{N} \sum_j \;\lgl \; S_{jf}^z \; \rgl \; .
\ee
The index $f=1$ corresponds to the ferroelectric (ordered) phase, while the index $f=2$, 
to the paraelectric (disordered) phase, so that
\be
\label{143}
 s_1 \; > \; s_2 \; ,
\ee  
and when the external strain is absent, then $s_1$ remains finite, but the order parameter
of the disordered phase tends to zero,
\be
\label{144}
s_2 \; \ra \; 0 \qquad ( D_j \ra 0 ) \;   .
\ee

For equal lattice sites, one can set
\be
\label{145}
\Om_j \; = \; \Om \; , \qquad  D_j = D_0  \; .
\ee
The value of $B_{ij}$ is usually small and can be omitted. The behavior of the system 
depends on the parameters
\be
\label{146}
u \; \equiv \; \frac{U}{J} \; , \qquad 
\om \; \equiv \; \frac{\Om}{J} \; , \qquad 
h \; \equiv \; \frac{D_0}{J} \; .
\ee
As is seen, the Hamiltonian (\ref{141}) is close by its structure to the Hamiltonian 
(\ref{137}). The overall behavior of the order parameters $s_f$ is analogous to that of
the order parameters of ferromagnets with mesoscopic paramagnetic fluctuations.

\subsection{Systems with mesoscopic density fluctuations}

Microscopic density fluctuations are characterized by phonons. But there also may happen 
mesoscopic density fluctuations so that the system represents an inhomogeneous two-density
fluctuating mixture. Systems with mesoscopic density fluctuations can model solids with 
regions of disorder or, more likely, liquid or gaseous systems with inhomogeneous density. 
As an example, let us consider a lattice-gas model 
\cite{Frenkel_5,Eyring_222,Cernuschi_223,Yang_224,Lee_225,Huang_226} modified so that to 
take into account mesoscopic density fluctuations 
\cite{Yukalov_227,Yukalov_228,Yukalov_229,Yukalov_230}.

The renormalized effective Hamiltonian has the form $H_{eff} = H_1 \bigoplus H_2$, where
the partial Hamiltonians are
$$
H_f \; = \; w_f \int \psi_f^\dgr(\br) \; \left[\; \hat H_L(\br) - 
\mu\; \right] \;\psi_f(\br) \; d\br \; +
$$
\be
\label{147}
+ \; 
\frac{1}{2} \; w_f^2 \int \psi_f^\dgr(\br) \; \psi_f^\dgr(\br') \;
\Phi(\br-\br') \; \psi_f(\br') \; \psi_(\br) \; d\br d\br' \; ,
\ee
here $\hat{H}_L(\br)$ is a Hamiltonian containing external fields, for instance a 
Hamiltonian of an optical lattice, and $\Phi({\bf r})$ is an interaction potential. The 
phases are distinguished by their density,
\be
\label{148}
 \rho_f \; \equiv \; \frac{N_f}{V_f} \; = \; 
\frac{1}{V} \int \lgl \; \psi_f^\dgr(\br) \; \psi_f(\br) \; \rgl \; d\br \; ,
\ee
with the number of atoms in phase $f$ being
\be
\label{149}
 N_f \; = \; 
w_f \int \lgl \; \psi_f^\dgr(\br) \; \psi_f(\br) \; \rgl \; d\br \;  .
\ee
One phase is denser than the other,
\be
\label{150}
  \rho_1 \; > \;  \rho_2 \; .
\ee
The lattice sites can be either occupied or free.

The field operators can be expanded over localized orbitals,
\be
\label{151}
\psi_f(\br) \; = \; \sum_{nj} e_{jf} \; c_{nj} \; \vp_{nj}(\br) \; ,
\ee
where the variable $e_{jf} = 0,1$, depending on whether the site is free or occupied, 
and the index $j = 1,2,\ldots,N_L$ enumerates the lattice sites. Each lattice site can 
host not more than one atom, which is expressed as the unipolarity condition
\be
\label{152}
 \sum_n c_{nj}^\dgr \;c_{nj} \; = \; 1 \qquad 
c_{mj}^\dgr \;c_{nj}^\dgr \; = \; 0 \;  .
\ee
The atoms filling the lattice are assumed to be well localized, such that the terms 
associated with the hopping of atoms from one lattice site to another are much smaller 
than the interaction terms of localized atoms. Then Hamiltonian (\ref{147}), in the 
single-band approximation, reduces to the lattice model
\be
\label{153}
H_f \; = \; - w_f \; \mu \sum_{j=1}^{N_L} e_{jf} +
\frac{1}{2} \; w_f^2 \sum_{i\neq j}^{N_L} \Phi_{ij} \; e_{if} \; e_{jf} \; .
\ee

The following calculations can either employ the variables $e_{fj} = 0,1$ or it can be 
convenient to pass to the pseudospin variables $S_{fj}^z = \pm 1/2$ through the canonical 
transformation
\be
\label{154}
 e_{jf} \; = \; \frac{1}{2} + S_{jf}^z \; , \qquad 
S_{jf}^z \; = \; e_{jf} - \; \frac{1}{2} \; .
\ee
In the latter case, we come to the pseudospin Hamiltonian
$$
H_f \; = \; \frac{1}{8} \; N_L \; \left( w_f^2 \Phi - 4 w_f \; \mu \right)  \; +
$$
\be
\label{155}
+ \; \frac{1}{2} \; 
\left( w_f^2 \Phi - 2 w_f \;\mu \right)  \sum_{j=1}^{N_L} S_{jf}^z + 
\frac{1}{2} \; 
w_f^2 \sum_{i\neq j}^{N_L} \Phi_{ij} \; S_{if}^z \; S_{jf}^z \; ,
\ee
in which
\be
\label{156}
 \Phi \; \equiv \; \frac{1}{N_L} \sum_{i\neq j}^{N_L} \Phi_{ij} \; .
\ee

The densities of the phases can be represented as 
\be
\label{157}
\rho_f \; = \; \frac{N_f}{V_f} \; = \; \frac{N_L}{2V} \; ( 1 + s_f ) \; ,
\ee
where the mean pseudospins are
\be
\label{158}
 s_f \; \equiv \; 2 \; \lgl \; S_{jf}^z \; \rgl \; .
\ee
Due to condition (\ref{150}), we have the inequality
\be
\label{159}
s_1 \; > \; s_2 \qquad ( \rho_1 > \rho_2) \; .
\ee

Similarly to other spin models with an external field, in the mean-field approximation 
the two-density mixed state with mesoscopic fluctuations can exist in a finite temperature 
range between the lower and upper nucleation temperatures. The other condition is the low 
filling factor $\nu$, such that
\be
\label{160}
 0 \; < \; \nu \; \equiv \; \frac{N}{N_L} \; < \; \frac{1}{2} \; .
\ee
Hence not more than half of the lattice sites can be occupied. This is why such a 
two-density fluctuating state corresponds rather to a gaseous substance. The lower 
nucleation temperature becomes zero for $\nu < 0.3$. The upper nucleation temperature 
$T_{nuc}^*$, where $w_1(T_{nuc}^*) = 0$, reads as
\be
\label{161}
 T_{nuc}^* \; = \; \frac{\nu \Phi}{(1-2\nu)\ln(1/\nu-1) } \;  .
\ee
It is useful to note that, as for other spin models, the mean-field approximation is not 
the best tool for describing spin-ordering effects in the disordered phase, such as 
paramagnetic phase. For more correctly characterizing the disordered phase, it is necessary 
to resort to more refined approximations taking into account short-range order, which, 
however, makes the consideration much more complicated.

\subsection{Mesoscopic structural fluctuations}

In the vicinity of a phase transition between two crystallographic structures with close 
properties there appear local structural fluctuations \cite{Michalski_231,Mitus_232}. 
Let us consider the mesoscopic fluctuations of one crystallographic structure inside
another structure. Following the general approach expounded above, the effective 
renormalized Hamiltonian reads as $H = H_1 \bigoplus H_2$, with $H_f$ for $f=1$ and 
$f=2$ corresponding to two different structures \cite{Yukalov_233,Yukalov_234}. This type
of fluctuations occurs in a variety of phase transition regions. Therefore we illustrate
the related method of description of such a typical situation with more detail.   

Starting with the standard form of the partial Hamiltonians (\ref{147}), where
\be
\label{162}
H_L(\br) \; = \; - \; \frac{\nabla^2}{2m} \; ,
\ee
we expand the field operators over the well-localized Wannier functions \cite{Marzari_235},
\be
\label{163}
\psi_f(\br) \; = \; \sum_j c_{jf} \; w(\br - \br_{jf}) \;   ,
\ee
keeping in mind the lowest band. The excited states will be considered through phonon 
excitations over the ground state \cite{Yukalov_236,Yukalov_237,Yukalov_238}. Since the
phase probabilities $w_f$ enter the Hamiltonian in a nontrivial way, we explain the 
derivation of the phonon Hamiltonian in sufficient details.  
 
In what follows, in the majority of expressions, not to overload notations, we omit the 
phase index $f$, keeping in mind that all quantities in the Hamiltonian $H_f$ depend 
on the type of the crystallographic structure, hence on the index $f$. Then Hamiltonian 
(\ref{147}) takes the form 
$$
H_f \; = \; - w_f \sum_{i\neq j} J_{ij} \; c_i^\dgr \; c_j + 
w_f \sum_j \left( \frac{p_j^2}{2m} - \mu \right) \; c_j^\dgr \; c_j \; +
$$
\be
\label{164}
+ \; 
\frac{1}{2} \; w_f^2 \sum_j U_{jj} \; c_j^\dgr \; c_j^\dgr \; c_j \; c_j +
\frac{1}{2} \; w_f^2 \sum_{i\neq j} U_{ij} \; c_i^\dgr \; c_j^\dgr \; c_j \; c_i \; ,   
\ee
in which
$$
J_{ij} \; = \; - 
\int w^*(\br-\br_i) \; \left( - \;\frac{\nabla^2}{2m}\right) \; w(\br-\br_j) \;d\br 
\qquad ( i \neq j) \; ,
$$
$$
U_{ij} \; = \; 
\int |\; w(\br-\br_i) \; |^2 \; \Phi(\br-\br') \; 
|\; w(\br'-\br_j)\; |^2 \; d\br d\br' \; ,
$$
\be
\label{165}
\bp_j^2 \; = \; 
\int w^*(\br-\br_j) \; ( -\nabla^2) \; w(\br-\br_j)\; \; d\br  \; .
\ee

There are two kinds of crystalline matter, being either a coherent periodic system, 
consisting of not localized particles \cite{Kirzhnits_240,Vozyakov_241} or a system 
of localized particles oscillating around their lattice sites. We consider a crystal 
formed by localized particles, whose oscillations are described as self-consistent
phonons \cite{Yukalov_36,Guyer_242}.    

We employ the no-double-occupancy constraint and no-hopping condition, according to 
which
\be
\label{166}
 c_j^\dgr \; c_j \; = \; 1 \; , \qquad c_j \; c_j \; = \; 0 \; ,
\qquad c_i^\dgr \; c_j \; = \; \dlt_{ij} \; .
\ee
The term containing the chemical potential can be omitted, since in the total Hamiltonian
$H$ it gives a constant term $w_1 \mu + w_2 \mu = \mu$. Then Hamiltonian (\ref{164}) 
simplifies to 
\be
\label{167}
 H_f \; = \; w_f \sum_j \frac{p_j^2}{2m} + 
\frac{1}{2} \; w_f^2 \sum_{i\neq j} U_{ij} \; .
\ee
 
The phase probabilities are defined as the minimizers of the thermodynamic potential, 
while respecting the normalization condition $w_1 + w_2 = 1$. This, with the notation 
\be
\label{168}
w_1 \; \equiv \; w \; , \qquad w_2 \; = \; 1 - w \; ,
\ee
gives the equation
\be
\label{169}
 \frac{\prt\Om}{\prt w} \; = \; 
\left\lgl \; \frac{\prt H}{\prt w} \; \right\rgl \; = \; 
\left\lgl \; \frac{\prt H_1}{\prt w} \; \right\rgl +
\left\lgl \; \frac{\prt H_2}{\prt w} \; \right\rgl \; = \; 0 \; .
\ee
From here, it follows
\be 
\label{170}
 w \; = \; \frac{2\Phi_2 + K_2 - K_1}{2(\Phi_1+\Phi_2)} \;  ,
\ee 
where, restoring the dependence on the phase index $f$, we define the kinetic and 
potential energies 
\be
\label{171}
K_f \; \equiv \; 
\left\lgl \; \frac{1}{N} \sum_j \frac{\bp_j^2}{2m} \; \right\rgl \; ,
\qquad 
\Phi_f \; \equiv \; 
\left\lgl \; \frac{1}{2N} \sum_{i\neq j} U_{ij}\; \right\rgl \;  .
\ee

Phonon excitations are introduced in the standard way 
\cite{Maradudin_243,Leibfried_244,Reissland_245}, by treating the spatial coordinate 
as an operator 
\be
\label{172}
\br_j \; = \; \ba_j + \bu_j
\ee
characterizing the deviation ${\bf u_j}$ from the lattice site ${\bf a_j}$, so that
\be
\label{173}
 \ba_j \; = \; \lgl \; \br_j \; \rgl \; , \qquad 
\lgl \; \bu_j \; \rgl \; = \; 0 \; .
\ee
Often, it is convenient to use the relative variables
\be
\label{174}
\br_{ij} \; \equiv \; \br_i - \br_j \; , \qquad 
\ba_{ij} \; \equiv \; \ba_i - \ba_j \; , \qquad 
\bu_{ij} \; \equiv \; \bu_i - \bu_j \; .
\ee

Recall that the effective particle interaction $U_{ij}$, defined in (\ref{165}), depends
on the type of the crystallographic lattice through the Wannier functions associated with
the considered crystalline phase, hence depending on the phase index $f$,
\be
\label{175}
U_{ij} \; = \; U_f(\br_{ij} ) \;   .
\ee
The potential energy is the sum
\be
\label{176}
B_f \; \equiv \; \frac{1}{2N} \sum_{i\neq j} U_f(\br_{ij} ) \; .
\ee
The latter can be expanded in powers of the deviations from the lattice sites $u_i^\al$
giving
\be
\label{177}
 B_f \; \simeq \; U_f + 
\frac{1}{2} \sum_{ij} \; \sum_{\al\bt} B_{ij}^{\al\bt} \; u_i^\al u_j^\bt \; ,
\ee
where the notation is used:
\be
\label{178}
U_f \; \equiv \; \frac{1}{2N} \sum_{i\neq j} U_f(\ba_{ij}) \; , 
\qquad
B_{ij}^{\al\bt} \; \equiv \; \frac{\prt^2 B_f}{\prt a_i^\al \prt a_j^\bt} \;  .
\ee
Then Hamiltonian (\ref{167}) becomes
\be
\label{179}
H_f \; = \;  w_f \sum_j \frac{\bp_j^2}{2m} + w_f^2 \; U_f N +
\frac{1}{2} \; w_f^2 
\sum_{ij} \; \sum_{\al\bt} B_{ij}^{\al\bt} \; u_i^\al u_j^\bt \; .
\ee

Introducing the phonon operators $b_{ks}$ by the transformations 
$$
\bu_j \; = \; \frac{1}{\sqrt{2N}} \sum_{ks} \frac{\bfe_{ks}}{\sqrt{m\om_{ks}} } \;
\left( b_{ks} + b_{-ks}^\dgr\right) \; e^{i\bk\cdot\ba_j} \; ,
$$
\be
\label{180}
\bp_j \; = \; -\; \frac{i}{\sqrt{2N}} \sum_{ks} \sqrt{m\om_{ks} } \; \bfe_{ks} \;
\left( b_{ks} - b_{-ks}^\dgr\right) \; e^{i\bk\cdot\ba_j} \; ,
\ee
and defining the phonon frequency $\omega_{ks}$ from the eigenproblem
\be
\label{181}
w_f \; \frac{1}{m} 
\sum_j \; \sum_\bt B_{ij}^{\al\bt} \; e^{i\bk\cdot\ba_{ij} } \; e_{ks}^\bt \; = \; 
\om_{ks}^2 \; e_{ks}^\al \; ,
\ee
where ${\bf e}_{ks}$ is a polarization vector, we come to the phonon Hamiltonian
\be
\label{182}
 H_f \; = \; w_f \sum_{ks} \; \left( b_{ks}^\dgr \; b_{ks} + \frac{1}{2} \right)
+ w_f^2 \; U_f \; N \;  .
\ee

To explicitly stress the dependence on the phase probability $w_f$, it is useful to define
the frequency
\be
\label{183}
\ep_{ks}^2 \; \equiv \; \frac{1}{m} \sum_j \; \sum_{\al\bt} 
B_{ij}^{\al\bt} \; e_{ks}^\al \; e_{ks}^\bt \;   e^{i\bk\cdot\ba_{ij}} \;  ,
\ee
such that
\be
\label{184}
 \om_{ks} \; = \; \sqrt{w_f} \; \ep_{ks} \;  .
\ee

It is straightforward to find the mean kinetic energy
\be
\label{185}
K_f \; = \; 
\frac{1}{4N} \sum_{ks} \om_{ks} \; \coth\left( \frac{w_f \om_{ks}}{2T} \right)
\ee
and the interaction energy
\be
\label{186}
 \Phi_f \; = \; \lgl \; B_f \; \rgl \; = \; U_f + 
\frac{1}{2N} \sum_{ij} \;
\sum_{\al\bt} B_{ij}^{\al\bt} \; \lgl \; u_i^\al u_j^\bt \; \rgl \;  ,
\ee
which are related by the equality
\be
\label{187}
w_f \; \Phi_f \; = \; w_f \; U_f + K_f \;   .
\ee
Then the kinetic energy can be written as 
\be
\label{188}
 K_f \; = \; w_f \; \frac{1}{2N} 
\sum_{ij} \; 
\sum_{\al\bt} B_{ij}^{\al\bt} \; \lgl \; u_i^\al u_j^\bt \; \rgl \; .
\ee
While for the phonon correlation function, we have
\be
\label{189}
\lgl \; u_i^\al u_j^\bt \; \rgl \;  = \; \frac{\dlt_{ij}}{2N} \sum_{ks} 
\frac{ e_{ks}^\al e_{ks}^\bt }{m\om_{ks}} \; \coth\left( \frac{w_f\om_{ks}}{2T}\right) \; .
\ee

The phase probability (\ref{170}) takes the form
\be
\label{190}
 w \; = \; \frac{2U_2+3(K_2-K_1)}{2(U_1+U_2)} \;  .
\ee
The probability $w = w_1$ shows the fraction of the system occupied by the crystallographic 
phase $f=1$ and, respectively, the fraction $w_2 = 1 - w$, of the phase $f=2$. Note that
the same expression (\ref{190}) can be obtained from the minimization of the free energy
$F = F_1 + F_2$ with respect to the phase probability $w$, where
\be
\label{191}
 F_f \; = \; w_f^2 \; U_f + 
\frac{T}{N} \sum_{ks} 
\ln \left[ \; 2\sinh\left( \frac{w_f\om_{ks}}{2T}\right) \; \right] \;  .
\ee

\subsection{Fluctuations in frustrated materials}

Structural fluctuations, considered in the previous section, can occur around the point
of structural phase transition. At sufficiently low temperature below the transition 
point the more stable structure prevails. However, there exist materials that do not
form a well defined crystalline structure even at very low temperatures because of the
frustration effect caused by competing interactions \cite{Bakai_289}. The spatial 
structure of such frustrated materials is formed of a combination of two or more 
coexisting structures and random regions. Frustrated materials are different from 
glass that is defined as ``a nonequilibrium, non-crystalline state of matter that appears 
solid on a short time scale but continuously relaxes towards the liquid state" 
\cite{Zanotto_290}. A frustrated material can possess a not completely ordered structure,
not relaxing to anything, even at its ground state \cite{Bakai_289}. The frustrated 
material also differs from the liquid state, where the particles are uniformly randomly
distributed in space, although can contain quasi-crystalline cluster fluctuations 
\cite{Ubbelohde_14,Hayes_15,Dash_16,Yukalov_246}. There exists a wide variety of
frustrated materials, such as complex polymers \cite{Ferreiro_292}.

With the evolution of nano-fabrication, it is now possible to create frustrated materials 
for realizing virtually any geometrical structures. This has recently opened a path 
toward the deliberate design of novel materials with exotic states often not found in 
natural substances \cite{Lookman_291}. In the present section, we consider a simple model 
of a frustrated matter that is formed of coexisting crystalline and random structures, 
where the coexistence is caused by competing interactions favoring different spatial 
orders.  

To describe the crystalline phase arising in a frustrated matter, we can use the theory 
of the previous section, where $w_1 \equiv w$ is the probability of the crystalline 
structure, while $w_2 = 1-w$ now will be the probability of a random structure. In order 
to stress the essence of the approach, we simplify the consideration by resorting to the 
Debye approximation. 

A very important quantity characterizing crystalline state is the phonon correlation 
function (\ref{187}) and the related mean-square deviation
\be
\label{192}
r_0^2 \; = \; \sum_{\al=1}^3 \; \lgl \; u_i^\al \; u_i^\bt \; \rgl \; ,
\ee
for which we have
\be
\label{193}
 r_0^2 \; = \; \frac{1}{2N} 
\sum_{ks} \frac{1}{m\om_{ks} } \; \coth\left( \frac{w\om_{ks}}{2T} \right) \;  .
\ee

Recall that $\omega_{ks} = \sqrt{w} \varepsilon_{ks}$. The latter can be written as
\be
\label{194}
\ep_{ks}^2 \; \equiv \; \sum_{\al\bt} D_k^{\al\bt} \; e_{ks}^\al \; e_{ks}^\bt \; ,
\ee
with the dynamical matrix
\be
\label{195}
 D_k^{\al\bt} \; \equiv \; 
\frac{1}{m} \sum_j B_{ij}^{\al\bt} \; e^{i\bk\cdot\ba_{ij} } \;  .
\ee
      
In the Debye approximation, one accepts the isotropic phonon spectrum defined as the 
average
\be
\label{196}
 \ep_k^2 \; \equiv \; \frac{1}{3} \sum_s \ep_{ks}^2 \; = \; 
\sum_\al D_k^{\al\al} \; \equiv \; D_k \;  ,
\ee
and the dynamical matrix is taken in the long-wave limit $D_k \simeq D k^2$, so that 
from (\ref{194}) one obtains the spectrum
\be
\label{197}
\ep_k \; = \; c_0 \; k \qquad ( 0 \leq k\leq k_D) \;   ,
\ee
where
\be
\label{198}
c_0 \; = \; \sqrt{D} \; , \qquad k_D \; = \; (6\pi^2\rho_1)^{1/3}    
\ee
and $\rho_1 = N_1/V_1$ is the average density.

Then the Debye approximation for the mean-square deviation reads as
\be
\label{199}
 r_0^2 \; = \; \frac{9w\rho_1}{2m\Theta_D\rho} 
\int_0^1 x\; \coth\left( \frac{\Theta_D}{2T}\; x\right) \; dx \;  ,
\ee
in which the effective Debye temperature
\be
\label{200}
\Theta_D \; \equiv \; w^{3/2} T_D
\ee
is renormalized by the presence of the random structure, as compared to the Debye 
temperature
\be
\label{201}
T_D \; \equiv \; c_0 k_D
\ee
of the pure crystal without such structural fluctuations. 

In stable crystals, the mean-square deviation is certainly smaller than the half of the 
distance between the nearest neighbors, $r_0^2 < a/2$, which is called the Lindemann
criterion of stability \cite{Lindemann_247}. Sometimes, one connects this criterion
with the melting of crystal. However this is exactly a stability criterion that is not 
directly connected with melting. The instability of crystals occurs at temperatures 
essentially higher the melting temperature, as confirmed by numerous calculations
\cite{Cotterill_32,Cotterill_34,Moleko_35,Yukalov_36,Haymet_248,Johnson_249,Zubov_250,
Zubov_251,Zubov_252,Zubov_253}. Also, the Lindemann instability can exist in optical 
lattices formed by laser beams, when the melting as such does not happen 
\cite{Yukalov_254,Yukalov_255}. A detailed review of the literature on melting is 
presented in Ref. \cite{With_256}.   

The free energy of the crystalline state, coexisting with the competing random structure, 
reads as
\be
\label{202}
 F_1 \; = \; w^2 U_1 + 9T \int_0^1 x^2 \; \ln\left[ \;
2\sinh\left( \frac{w^{3/2}T_D}{2T}\; x\right) \; \right] \; dx \;  .
\ee

For the random-structure phase, the grand Hamiltonian 
$$
H_2 \; = \; w_2 \int \psi_2^\dgr(\br) 
\left(-\; \frac{\nabla^2}{2m} - \mu\right) \psi_2(\br) \; d\br \; +
$$
\be
\label{203}
 + \;
\frac{1}{2} \; w_2^2 \int \psi_2^\dgr(\br) \; \psi_2^\dgr(\br') \; \Phi(\br-\br') \;
\psi_2(\br') \; \psi_2(\br) \; d\br d\br' 
\ee
has to be complimented with an appropriate approximation. Typical particle interactions
usually contain nonintegrable hard cores precluding from the use of the Hartree or 
Hartree-Fock approximations due to the nonintegrability of interaction potentials. This 
problem can be cured by resorting to the Kirkwood approximation \cite{Kirkwood_192} that 
is a correlated Hartree approximation
$$
\psi_2^\dgr(\br) \; \psi_2^\dgr(\br') \; \psi_2(\br') \; \psi_2(\br) \; = \;
g(\br-\br') \left\{ \;
 \psi_2^\dgr(\br) \; \psi_2(\br) \lgl \; \psi_2^\dgr(\br') \; \psi_2(\br') \; \rgl + \right.
$$
\be
\label{204}
+ \left.
 \lgl \; \psi_2^\dgr(\br) \; \psi_2(\br) \; \rgl \psi_2^\dgr(\br') \; \psi_2(\br') -
 \lgl \; \psi_2^\dgr(\br) \; \psi_2(\br) \; \rgl 
\lgl \; \psi_2^\dgr(\br') \; \psi_2(\br') \; \rgl \;
\right\} \; ,
\ee
where $g({\bf r})$ is a pair correlation function. Notice that averaging the relation 
(\ref{204}) yields the definition of the pair correlation function
\be
\label{205}
 g(\br-\br') \; = \; 
\frac{\lgl\;\psi_2^\dgr(\br)\;\psi_2^\dgr(\br')\;\psi_2(\br')\;\psi_2(\br)\;\rgl }
{\lgl\;\psi_2^\dgr(\br)\;\psi_2(\br)\;\rgl \lgl\;\psi_2^\dgr(\br')\;\psi_2(\br')\;\rgl}\; .
\ee

It has been shown \cite{Yukalov_107,Yukalov_257} that, starting with the Kirkwood 
approximation, it is possible to develop a self-consistent perturbation theory for Green
functions of any order. In this theory, the interaction potential always enters as the 
regularized potential
\be
\label{206}
\overline\Phi(\br) \; \equiv \; g(\br) \; \Phi(\br)   
\ee
that is always integrable. A convenient way for defining the regularizing correlation 
function $g(r)$ is given in \cite{Yukalov_2025}.

The average density of the random phase
\be
\label{207}
\rho_2 \; \equiv \; \frac{N_2}{V_2} \; = \; \frac{1}{V}
\int  \lgl \; \psi_2^\dgr(\br) \; \psi_2(\br) \; \rgl d\br      
\ee
defines the chemical potential $\mu$. Keeping in mind its randomness, the density of 
the random phase can be taken as uniform,
\be
\label{208}
  \lgl \; \psi_2^\dgr(\br) \; \psi_2(\br) \; \rgl \; = \; \rho_2 \;  .
\ee

In the Kirkwood approximation, the grand Hamiltonian (\ref{203}) becomes
\be
\label{209}
 H_2 \; = \; w_2 \int \psi_2^\dgr(\br) \; \left( - \; \frac{\nabla^2}{2m} + 
w_2 \rho_2 \Phi_0 - \mu \right) \; \psi_2(\br) \; d\br + E_0 \;  ,
\ee
where
\be
\label{210}
\Phi_0 \; \equiv \; \int \overline \Phi(\br) \; d\br
\ee
and 
\be
\label{211}
E_0 \; = \; - \frac{1}{2} \; w_2^2 \; \frac{\rho_2^2}{\rho} \; \Phi_0 \; N \;   .
\ee

Passing to the Fourier expansion for the field operators, we have
\be
\label{212}
 H_2 \; = \; w_2 \sum_k \om_k \; a_k^\dgr \; a_k + E_0 \;  ,
\ee
with the spectrum
\be
\label{213}
\om_k \; = \; \frac{k^2}{2m} + w_2 \rho_2 \Phi_0 - \mu \;  .
\ee

The grand potential per particle 
$$
\Om_2 \; = \; - \frac{T}{N} \; \ln{\rm Tr} \; e^{-\bt H_2}
$$
reduces to 
\be
\label{214}
\Om_2 \; = \; \pm T \sum_k \ln\left( 1 \mp e^{-\bt w_2 \om_k} \right) + E_0 \;  .
\ee
Assuming that $\exp{\beta \mu_{eff}} \ll 1$ yields 
\be
\label{215}
\Om_2 \; = \; - \; \frac{T}{\rho} \; 
\left( \frac{m T}{2\pi w_2}\right)^{3/2} \; e^{\bt\mu_{eff}} + E_0 \; ,
\ee
where 
\be
\label{216}
\mu_{eff} \; \equiv \; w_2 \; ( \mu - w_2 \rho_2 \Phi_0 ) \;  .
\ee
The density of the random phase (\ref{207}) takes the form
\be
\label{217}
 \rho_2 \; = \; \left( \frac{ mT}{2\pi w_2}\right)^{3/2} \; e^{\bt\mu_{eff}} \;  ,
\ee
which leads to
\be
\label{218}
w_2\; \mu \; = \; w_2^2 \; \rho_2 \; \Phi_0 + 
\frac{3}{2} \; T \; \ln\left( 4\pi \; w_2 \; \frac{E_k}{T}\right) \; ,
\ee
with the notation for the characteristic kinetic energy
\be
\label{219}
 E_k \; \equiv \; \frac{\rho^{3/2}}{2m} \; \cong \; 
\frac{\hbar^2}{2ma^2} \;  .
\ee 
Thus the grand potential per particle for the random phase is
\be
\label{220}
 \Om_2 \; = \; - T \; \frac{\rho_2}{\rho} + \frac{E_0}{N} \;  .
\ee
 
To simplify the equations, we can keep in mind that the random and crystalline phases 
differ in spatial structure, but do not differ much in the average densities, so that 
$\rho_1 \approx \rho_2 \approx \rho$. For the free energy of the random phase containing 
clusters of crystalline fluctuations, $F_2 = \Omega_2 + w_2 \mu$, we obtain 
\be
\label{221}
F_2 \; = \; \frac{1}{2} \; w_2^2 \; \rho \; \Phi_0 + 
\frac{3}{2} \; T \; \ln\left( 4\pi \; w_2 \; \frac{E_k}{T}\right) - T \;   .
\ee
The total free energy of the system is $F = F_1 + F_2$, with the phase probabilities 
$w = w_1$ and $w_2 = 1 - w$ defined as the minimizers of the free energy $F$.  

For the convenience of numerical calculations, let us introduce the following 
dimensionless parameters: the relative cohesive energy
\be
\label{222}
 u \; \equiv \;  \frac{U_1}{T_D} \; ,
\ee
the ratio of the characteristic potential energy to the Debye temperature
\be
\label{223}
 g \; \equiv \;  \frac{\rho\Phi_0}{2T_D} \; ,
\ee
and the ratio of the characteristic kinetic energy to the Debye temperature
\be
\label{224}
 e_K \; \equiv \;  \frac{E_K}{T_D} \;  .
\ee

The dimensionless free energy of the system can be written in the form
\be
\label{225}
 f \; = \; f_1 + f_2 \;  ,
\ee
in which
\be
\label{226}
  f \; \equiv \;  \frac{F}{T_D} \; , \qquad 
f_1 \; \equiv \;  \frac{F_1}{T_D} \; , \qquad 
f_2 \; \equiv \;  \frac{F_2}{T_D} \; .
\ee

Measuring temperature in units of the Debye temperature, we have 
$$
f_1 \; = \; w^2 \; u + 9T \int_0^1 x^2 \; 
\ln\left[ \; 2\sinh\left( \frac{w^{3/2}}{2T}\; x\right) \; \right] \; dx \; ,
$$
\be
\label{227}
f_2 \; = \; (1 - w)^2 \; g + 
\frac{3}{2} \; T \; \ln \left[ \; 4\pi (1 - w) \; \frac{e_k}{T} \; \right] - T \; .
\ee
The probability of the crystalline phase $w$ is defined as a minimizer of the free energy:
\be
\label{228}
  \frac{\prt f}{\prt w} \; = \; 0 \; , \qquad 
\frac{\prt^2 f}{\prt w^2} \; > \; 0 \; .
\ee

The free energy $f$ of the coexisting frustrated system, combining the crystalline phase
and the random uniform phase, has to be compared with the free energies of the pure 
crystalline solid, 
\be
\label{229}
  f_{sol} \; = \; u + 9T \int_0^1 x^2 \; 
\ln \left[\; 2\sinh\left( \frac{x}{2T}\right) \; \right] \; dx \; ,   
\ee
and of pure random structure,
\be
\label{230}
 f_{ran} \; = \; g + 
\frac{3}{2} \; T \; \ln\left( 4\pi \; \frac{e_k}{T}\right) - T \;  .
\ee
    
The general behavior of concrete frustrated materials, modelled by the competing 
crystalline and random structures, can be done by fixing the parameters $u$, $g$, 
and $e_K$. These parameters for many ordinary solids can easily be found in the books 
on solid-state physics \cite{Guyer_242,Pollack_258,Ashcroft_259,Kittel_260}. However, 
frustrated materials are less studied and their parameters are less known. In addition, 
employing the advance of nano-fabrication, it is nowadays possible to form a variety 
of novel frustrated materials with different parameters often not found in natural 
substances \cite{Lookman_291}. The typical behavior of the free energies for the 
considered frustrated system at low temperature below the Debye temperature $(T \ll 1)$ 
is shown in Fig. 1. We keep in mind that the cohesive parameter $u$ has to be negative, 
characterizing the potential well for a particle, with the absolute value comparable 
with the Debye temperature, while the relative kinetic parameter $e_K$ is positive 
and small, being of order $10^{-3}$. The frustration parameter $g$ is usually positive 
\cite{Bakai_289}.  

\begin{figure}[ht]
\centerline{
\hbox{ \includegraphics[width=8cm]{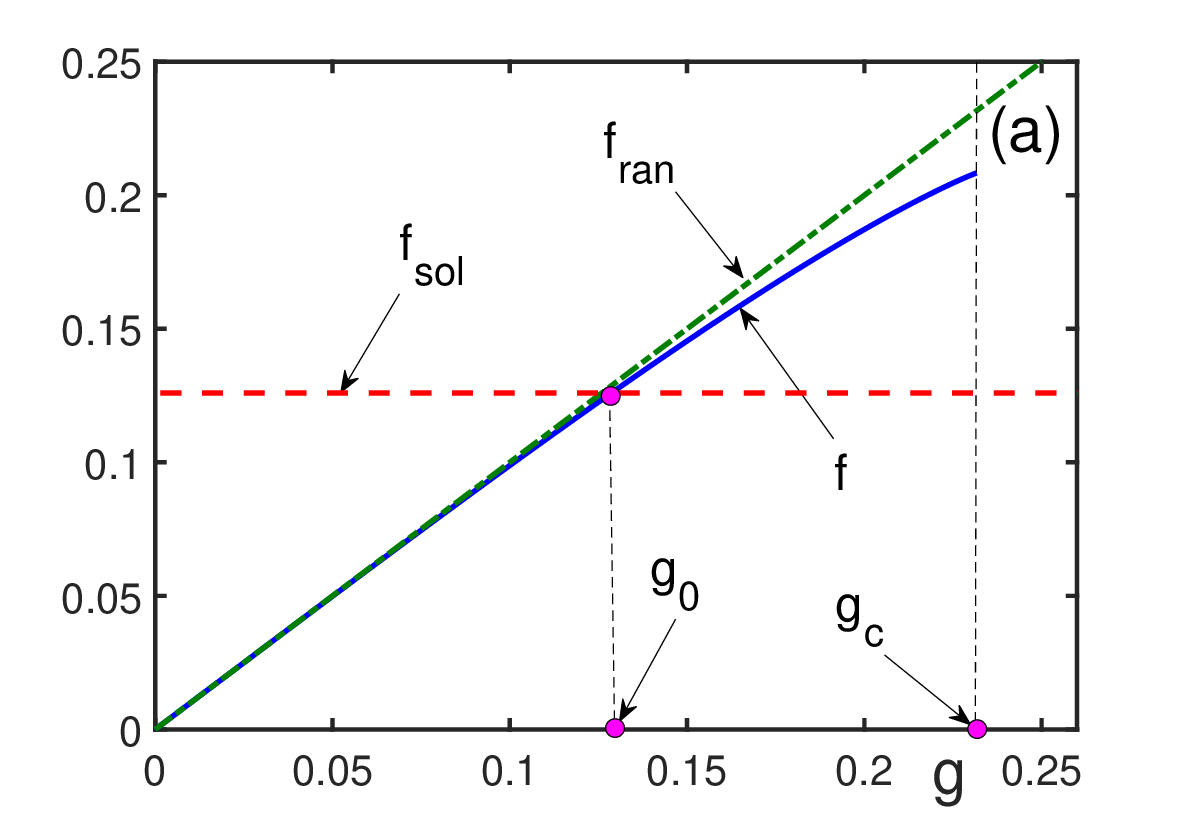}\hspace{1cm}
\includegraphics[width=8cm]{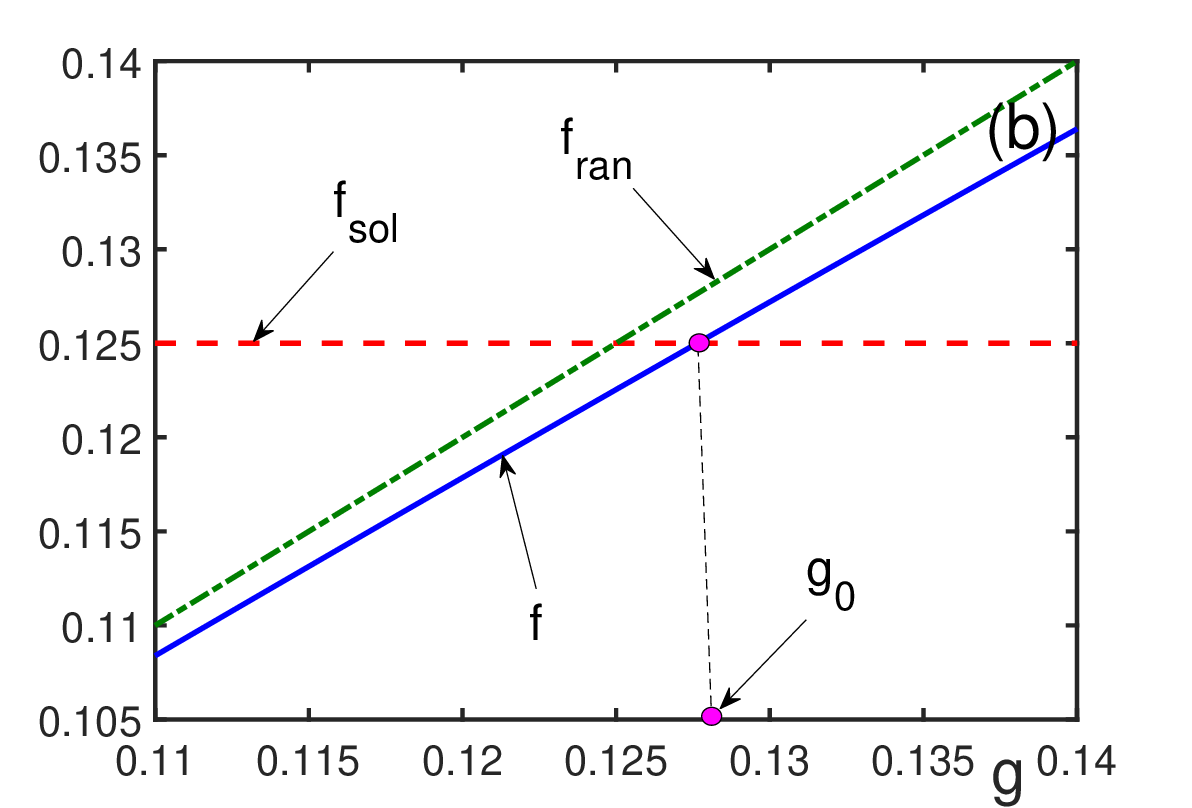}  } }
\caption{\small 
Free energies as functions of the frustration parameter $g$ under the fixed
cohesive-energy parameter $u = -1$: (a) Free energy of the frustrated system $f$ 
(solid line), compared with the free energies of a pure crystalline solid $f_{sol}$ 
(dashed line) and a pure random matter $f_{ran}$ (dashed-dotted line); (b) Detalized
region of the frustration parameter $g$, where the free energy of the frustrated matter 
crosses the free energy of the crystalline solid. 
}
\label{fig:Fig.1}
\end{figure}

By varying the frustration parameter $g$ from zero to $g_0$, we see that the frustrated 
system has the lowest free energy up to the value $g_0$, where it transforms into a pure
crystalline solid by a first order phase transition. The related probability of the 
crystalline phase is shown in Fig. 2. This behavior demonstrates that for the frustration 
parameter between $0$ and $g_0$ the frustrated structure is preferable, but it becomes
thermodynamically unfavorable for too strong frustration above $g_0$.  

\begin{figure}[ht]
\centerline{
\includegraphics[width=10cm]{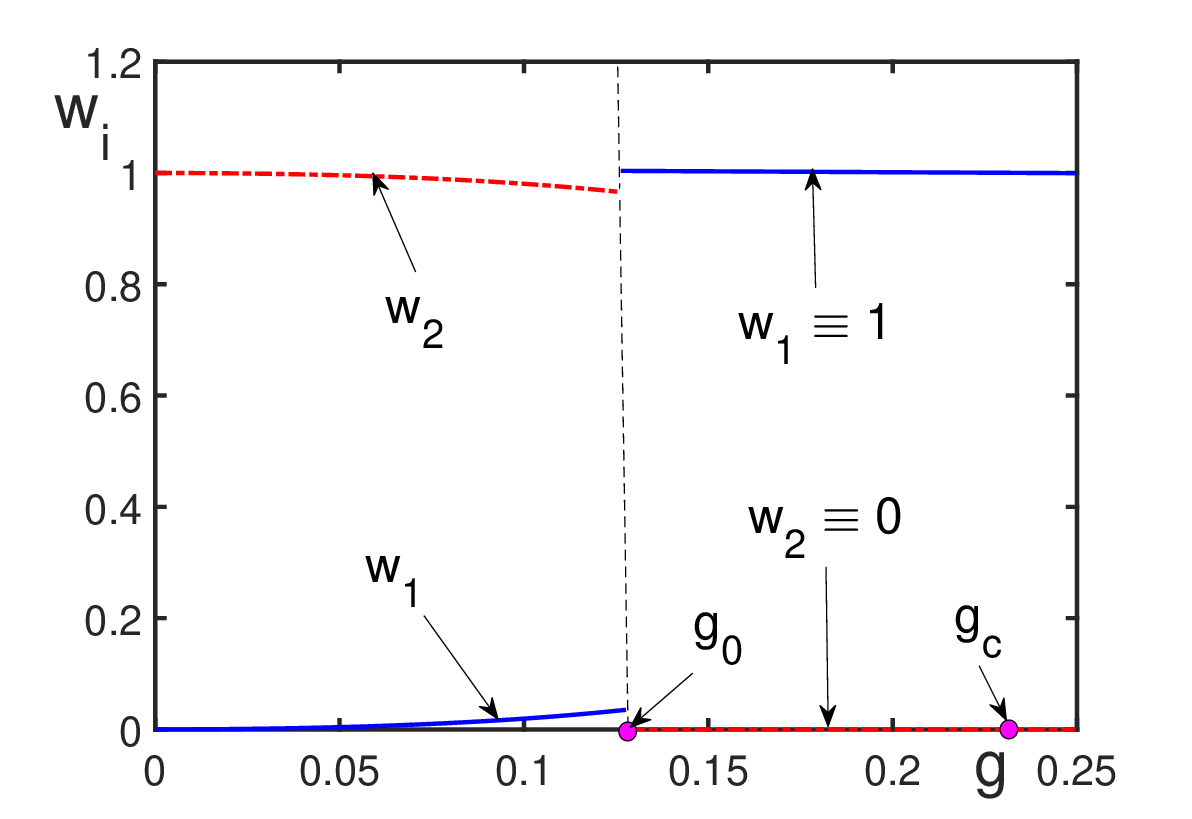}
 }
\caption{\small
The probability of the crystalline $w_1$ (solid line) and random $w_2$ 
(dashed line) fractions in a frustrated material as functions of the frustration 
parameter $g$, under fixed $u = -1$. The point of the phase transition between the 
frustrated matter and crystalline solid is $g_0$ = 0.128; the upper bound for the 
existence of the frustrated matter is $g_c=0.231632$. 
}
\label{fig:Fig.2}
\end{figure}

\subsection{Superfluid fluctuations in solids}

If in a solid there can appear regions of disorder, such as pores and dislocations, the
question arises whether these regions can house the embryos of liquid-like phase, and 
if so, whether these liquid-like formations could exhibit the property of superfluidity 
and the related Bose condensation. The possibility of arising superfluid properties in 
solids with dislocations, e.g. in solid $^4$He, has been recently widely discussed, as 
can be inferred from the reviews \cite{Prokofev_261,Boninsegni_262,Chan_263,Kuklov_264,
Yukalov_265,Fil_266}. However, no convincing experimental evidence for the existence 
of superfluidity in the dislocations of solid $^4$He has been presented. Although some 
Monte Carlo calculations do show the occurrence of superfluidity in solid $^4$He with 
dislocations (see the above reviews), however it is not possible to rule out that the 
signs of superfluidity are caused by finite-size and finite-time effects. The recent 
rather accurate numerical calculations employing path-integral ground state method 
\cite{Koning_267} provide compelling evidence for the absence of intrinsic superfluidity 
in dislocation cores of hcp $^4$He.

In order to find out whether liquid-like mesoscopic fluctuations in a solid with disorder 
regions could exhibit superfluid properties, it is necessary to study an explicit model 
of such a solid. A statistical model of a solid with mesoscopic liquid-like fluctuations 
allowing for the appearance of superfluidity has been considered in Ref. \cite{Yukalov_268}.

Let the field operators $\psi_1(\br)$ correspond to solid phase and $\psi_2(\br)$,
to liquid-like phase. The density of the solid phase is periodic over the lattice vectors,
\be
\label{231}
 \lgl \; \psi_1^\dgr(\br+\ba) \; \psi_1(\br+\ba) \; \rgl \; = \;
 \lgl \; \psi_1^\dgr(\br) \; \psi_1(\br) \; \rgl \;  ,
\ee
while the liquid phase is uniform,
\be
\label{232}
\lgl \; \psi_2^\dgr(\br) \; \psi_2(\br) \; \rgl \; = \;
 \lgl \; \psi_2^\dgr(0) \; \psi_2(0) \; \rgl \;  .
\ee

The liquid phase allows for the global gauge symmetry breaking,
\be
\label{233}
 \lgl \; \psi_2^\dgr(\br)  \; \rgl \; = \; \sqrt{\rho_0} \; \neq \; 0
\qquad
\left( \rho_0 \equiv \frac{N_0}{V_2} \right) \; ,
\ee
which is necessary for the occurrence of Bose condensate, with density $\rho_0$. Note that
the global gauge symmetry breaking is a necessary and sufficient condition for Bose-Einstein
condensation \cite{Yukalov_111,Yukalov_114,Lieb_269,Lieb_270}. The latter is a sufficient 
condition for superfluidity. The symmetry breaking is realized by the Bogolubov 
\cite{Bogolubov_171,Bogolubov_172,Bogolubov_173} shift 
\be
\label{234}
 \psi_2(\br) \; = \; \eta(\br) + \psi_{un}(\br) \;  ,
\ee
in which 
\be
\label{235}
\eta(\br) \; = \; \lgl \; \psi_2(\br) \; \rgl \; = \; \sqrt{\rho_0}
\ee
is the order parameter or condensate wave function and $\psi_{un}({\bf r})$ is the field 
operator of uncondensed particles. The fractions of condensed and uncondensed particles 
are denoted as
\be
\label{236}
n_0 \; \equiv \; \frac{N_0}{N_2} \; , \qquad
n_{un} \; \equiv \; \frac{N_{un}}{N_2} \;    ,
\ee
where $N_0$ is the number of condensed particles, $N_{un}$ is the number of uncondensed 
particles, and $N_2$ is the number of particles in the liquid phase.    
 
The gauge symmetry breaking leads to the nonzero anomalous density
\be
\label{237}
 \sgm(\br) \;  \equiv \; 
\frac{1}{\rho} \; \lgl \; \psi_{un}(\br) \; \psi_{un}(\br) \; \rgl \; = \; \sgm
\ee
that principally cannot be neglected as soon as the symmetry is broken. The anomalous 
average, together with condensed particles, define the sound velocity $s$ by the equation
\be
\label{238}
 s^2 \; = \; 4\pi \gm \; (n_0 + \sgm ) \;  .
\ee

The crystalline phase with liquid-like fluctuations is described as is explained in the 
previous section. Keeping in mind the zero-temperature case, we get the energy
\be
\label{239}
E_1 \; = \; w_1^2 \; U_1 + \frac{9}{8}\; w_1^{3/2} \; T_D \qquad ( T = 0 ) \;   ,
\ee
with the cohesive or configurational energy
$$
 U_1 \; \equiv \; \frac{1}{2N} \sum_{i\neq j} U_1(\ba_{ij} ) \;  .
$$

The superfluid Bose-condensed phase is described by the self-consistent approach advanced 
in Refs. \cite{Yukalov_271,Yukalov_272,Yukalov_273,Yukalov_274,Yukalov_275,Yukalov_276} 
(see also \cite{Yukalov_111,Yukalov_114,Yukalov_115,Yukalov_119,Yukalov_121,Yukalov_185,
Yukalov_186}). For the energy at zero temperature, we have
\be
\label{240}
 \frac{E_2}{E_K} \; = \; \frac{16 s^5}{15\pi^2} \; w_2^{7/2} + 4\pi \gm w_2^2 \;
\left( 1 + n_{un}^2 - 2n_{un} \; \sgm - \sgm^2 \right) \;  .
\ee
The phase probabilities are defined as the minimizers of the total system energy in the 
presence of mesoscopic fluctuations,
\be
\label{241}
 E \; = \; E_1(w_1) + E_2(w_2) \; .
\ee
This energy has to be compared with the energy $E_{sol}(1)$ of the pure solid phase and the 
energy $E_{liq}(1)$ of the pure liquid phase. At zero temperature, the most stable state of 
the system is defined by the lowest energy. It is shown \cite{Yukalov_268} that, in general, 
there could exist crystals with superfluid fluctuations, arising inside dislocations, 
provided materials with appropriate parameters would exist. 

It is interesting to consider the particular case of hcp $^4$He, whose characteristic 
parameters have been widely studied 
\cite{Hodgdon_277,Maris_278,Cazorla_279,Vitiello_280,Casorla_281,Chan_282}. Our 
investigation \cite{Yukalov_268} demonstrates that, in the frame of the considered model, 
hcp $^4$He cannot exhibit superfluid properties, in agreement with numerical studies 
\cite{Koning_267}.

\subsection{Superconductors with mesoscopic fluctuations}

As is mentioned in Sec. 2.6, the coexistence of superconducting and non-superconducting 
(normal-phase) fluctuations was predicted \cite{Shumovsky_82,Yukalov_83} yet before 
high-temperature superconductors were discovered in experiments \cite{Bednorz_84}. The 
model was generalized and studied for superconductors with an isotropic gap 
\cite{Yukalov_85,Coleman_86} and anisotropic gap \cite{Yukalov_87,Yukalov_88}. 
The mesoscopic phase separation into a hole-rich superconducting phase and a hole-poor
insulating antiferromagnetic phase has been confirmed in numerous theoretical and 
experimental works \cite{Phillips_89,Benedek_90,Korzhenevskii_91,Muller_92,Sigmund_93,
Nagaev_94,Wubbeler_95,Balagurov_96,Balagurov_97,Pomjakushin_98,Bianconi_99,Kivelson_100}.
This mesoscopic phase separation in superconducting materials is a particular type of 
self organization into the regions called {\it superstripes} that are the bubbles of 
$100-300$ \AA, and on a short time scale of the order of $10^{-12}$ s.

The field operators of charged Fermi particles in phase $f$, with spin $s$, are denoted 
as $\psi_{sf}({\bf r})$. The phases are distinguished by order parameters or order indices
\cite{Coleman_165,Coleman_166,Coleman_167,Coleman_168,Yukalov_169}. Let the superconducting 
phase be labeled by $f=1$ and non-superconducting, by $f=2$. The order parameter for the
superconducting phase is the non-zero anomalous average
\be
\label{242}
\lgl \; \psi_{s1}(\br) \; \psi_{-s1}(\br) \; \rgl \; \not\equiv \; 0 \;   ,
\ee
while for the non-superconducting phase the anomalous average is zero,
\be
\label{243}
\lgl \; \psi_{s2}(\br) \; \psi_{-s2}(\br) \; \rgl \; \equiv \; 0 \;   .
\ee
The field operators can be expanded over a complete set of functions,
\be
\label{244}
\psi_{sf}(\br) \; = \; \sum_k c_{sf}(\bk) \; \vp_k(\br)
\ee
that are the plane waves for a uniform system, ${\bf k}$ being momentum, or Bloch functions 
for a crystalline lattice, ${\bf k}$ being quasi-momentum.      

In the momentum representation, the order parameters are the anomalous averages
\be
\label{245}
 \sgm_f \; \equiv \; \lgl \; c_{-sf}(-\bk) \; c_{sf}(\bk) \; \rgl \; ,
\ee
satisfying the conditions 
\be
\label{246}
\sgm_1(\bk) \; \not\equiv \; 0 \; , \qquad \sgm_2(\bk) \; \equiv \; 0 \;   .
\ee
It is possible to work in a restricted space where spin and momentum are explicitly conserved,
so that
$$
c_{sf}^\dgr(\bk) \; c_{zf}(\bp) \; = \; 
\dlt_{sz} \; \dlt_{kp} \; c_{sf}^\dgr(\bk) \; c_{sf}(\bk) \; ,
$$
\be
\label{247}
c_{sf}(\bk) \; c_{zf}(\bp) \; = \; 
\dlt_{-sz} \; \dlt_{-kp} \; c_{-sf}(-\bk) \; c_{sf}(\bk) \; .
\ee

The system Hamiltonian is $H_{eff} = H_1 \bigoplus H_2$, where
$$
H_f \; = \; w_f \sum_s \int \psi_{sf}^\dgr(\br) \; \left[\; \hat K_f(\br) - \mu \; \right] \;
\psi_{sf}(\br) \; d\br \; + 
$$
\be
\label{248}
+ \;
\frac{1}{2} \; w_f^2 \sum_{sz} \int \psi_{sf}^\dgr(\br) \; \psi_{zf}^\dgr(\br') \; 
V_f(\br,\br') \; \psi_{zf}(\br') \; \psi_{sf}(\br) \; d\br d\br' \; .
\ee
Here $\hat{K}_f({\bf r})$ is a kinetic term and $V_f({\bf r}, {\bf r}')$ is an effective 
potential of particle interactions. 

Resorting to the Hartree-Fock-Bogolubov approximation, it is straightforward to diagonalize 
the Hamiltonians $H_f$ by means of the Bogolubov canonical transformation
\be
\label{249}
 c_{sf}(\bk) \; = \; u_f(\bk) \; a_{sf}(\bk) + v_f(\bk) \; a_{-sf}^\dgr(\bk) \; ,
\ee
in which the coefficient functions $u_f({\bf k})$ and $v_f({\bf k})$ are defined by the 
diagonalization condition. 
 
One more form of the order parameters is the gap
\be
\label{250}
\Dlt_f(\bk) \; = \; w_f \sum_p J_f(\bk,\bp) \; \sgm_f(\bp) \; ,
\ee
where $J_f({\bf k}, {\bf p})$ is a matrix element of the effective interaction potential. 
Since the gap is connected with the anomalous average, for the superconducting and normal 
phases we have
\be
\label{251}
  \Dlt_1(\bk) \; \not\equiv \; 0 \; , \qquad \Dlt_2(\bk) \; \equiv \; 0 \; .
\ee

Detailed explanations of calculations are given in Refs. 
\cite{Yukalov_83,Yukalov_85,Coleman_86,Yukalov_87}. For the isotropic gap, depending only 
on the modulus of the momentum, at the Fermi surface, the gap $\Delta \equiv \Dlt_1(k_F)$, 
where $k_F \approx (3 \pi^2 \rho_e)^{1/3}$, and $\rho_e$ is the charged particles density, 
the gap equation is
\be
\label{252}
\int_0^1 \frac{\Lbd}{ \sqrt{(\Dlt/\sqrt{w_1}\; \om_0)^2+ x^2 } } \;
\tanh\left[\; 
\frac{w_1^{3/2}\; \sqrt{(\Dlt/\sqrt{w_1}\; \om_0)^2+x^2}}{2T}\; \om_0\; \right] \; dx \; 
= \; 1 \; ,
\ee
where $\omega_0$ is the characteristic phonon frequency and
\be
\label{253}
\Lbd \; \equiv \; w_1 \; N(0) \; \left[ \; 
\frac{|\;\al\;|^2}{w_1\om_0^2} - J_C(k_F) \; \right]
\ee
is the effective coupling parameter. Here $N(0)$ is the density of states at the Fermi 
surface. The first term of the coupling parameter is due to electron-phonon coupling with 
the parameter $\alpha$. The second term 
\be
\label{254}
J_C(k_F) \; = \; 
\frac{\pi e_0^2}{k_F^2} \; \ln \; \left( 1 + 4 \; \frac{k_F^2}{\varkappa^2} \right)
\ee
is caused by Coulomb interactions, with $e_0$ being the carrier charge and the screening 
radius $\varkappa^{-1}$ defined by
$$
\varkappa^2 \; = \; \frac{4}{a_B} \; \left( \frac{3}{\pi}\; \rho_e \right)^{1/3} \;  .
$$ 

Introducing the notation for the dimensionless electron-phonon coupling parameter 
\be
\label{255}
\lbd \; \equiv \; N(0) \; \frac{|\;\al\;|^2}{\om_0^2}
\ee
and for the Coulomb coupling 
\be
\label{256}
\mu^* \; \equiv \; N(0) \; J_C(k_F)
\ee
allows us to write the effective coupling (\ref{253}) as
\be
\label{257}
 \Lbd \; = \; \lbd - w_1 \; \mu^* \;  .
\ee

The equation for the critical temperature follows from (\ref{252}) by setting zero gap, which
gives
\be
\label{258}
(\lbd - w_1 \; \mu^* ) \int_0^1 \frac{1}{x} \; 
\tanh\left( \frac{w_1^{3/2}\om_0}{2T_c} \; x \right) \; dx \; = \; 1 \;   .
\ee
As is evident, the superconductivity can exist provided that $\lambda > w_1 \mu^*$. Hence
superconductivity can occur in bad conductors for which $\lambda < \mu^*$, so that 
superconductivity could not arise in a pure sample, but the appearance of mesoscopic phase 
fluctuations reduces the influence of the Coulomb interaction and superconductivity becomes 
possible.   

The consideration can be extended to the case of anisotropic superconductors whose properties 
depend on the direction of momentum \cite{Yukalov_87}. In that case, the effective interaction
is expended over a basis characterizing the system anisotropy
\be
\label{259}
 J_1(\bk,\bp) \; = \; \sum_{ij} J_{ij} \; \chi_i(\bk) \; \chi_j^*(\bp) \;  .
\ee
The gap is also expanded over this basis,
\be
\label{260}
 \Dlt(\bk) \; = \; \sum_i \Dlt_i \; \chi_i(\bk) \;  ,
\ee
which reduces the gap equation to the system of uniform algebraic equations of the type
\be
\label{261}
 \Dlt_i \; = \; \sum_j A_{ij} \; \Dlt_j \;  .
\ee
The system of uniform algebraic equations enjoys nontrivial solutions when
\be
\label{262}
 {\rm det} ( \hat 1 - \hat A ) \; = \; 0 \;  ,
\ee
where $\hat{A}$ is the matrix of elements $A_{ij}$.

The analysis shows that superconductors with mesoscopic fluctuations enjoy the following 
properties distinguishing these superconductors from those without such fluctuations:
\begin{enumerate}
\item
Mesoscopic fluctuations in a superconductor can arise only when, in addition to effective 
attracting electron-phonon interactions there exist repulsive Coulomb interactions.

\item
Mesoscopic fluctuations enable the appearance of superconductivity in a sample even if
it were impossible in the matter without such fluctuations, for instance in bad conductors.

\item
The critical temperature for a mixture of gap waves with different symmetry is higher 
than the critical temperature related to any pure gap wave from this mixture, so that the
mixture of $s$-wave and $d$-wave superconductivity enjoys a higher temperature than any of 
these waves separately. 

\item
In bad conductors, the critical temperature as a function of the superconducting fraction 
has the typical bell shape \cite{Yukalov_87}.
\end{enumerate}

\section{Debye-Waller and M\"{o}ssbauer factors}

The appearance of mesoscopic fluctuations in the vicinity of phase transition points can 
be noticed, e.g. by  measuring sound velocity. In the presence of mesoscopic fluctuations
of two coexisting phases (see Secs. 6.4 and 6.5), sound velocity has the form
\be
\label{263}
 s \; = \; w_1^{3/2} \; c_1 + w_2^{3/2} \; c_2 \;  ,
\ee
where $c_f$ is the sound velocity in a pure phase $f$. At the phase transition point, the 
sound velocity experiences a noticeable decrease because of scattering on mesoscopic 
fluctuations. This decrease can be characterized by the relative sagging
\be
\label{264}
  \dlt_s \; = \; \frac{s-c_f}{c_f} \; .
\ee
If the sound velocity in pure phases does not vary much, so that $c_1 \approx c_2$, then,
assuming that $w \approx 1/2$, we find $\delta_s \approx -0.29$.  

As far as the existence of mesoscopic fluctuations influences the mean-square deviation 
of particles, the appearance of these fluctuations can be revealed by measuring the 
M\"{o}ssbauer or Debye-Waller factors. Since mesoscopic fluctuations are the most 
pronounced in the vicinity of phase transitions, where they lead to the mean-square 
deviation increase, the M\"{o}ssbauer and Debye-Waller factors should exhibit noticeable 
sagging in the vicinity of transition points
\cite{Yukalov_64,Yukalov_202,Yukalov_233,Yukalov_234,Yukalov_283,Yukalov_284,Yukalov_285}.

The total Debye-Waller or Lamb-M\"{o}ssbauer factor for a mixture of two phases is the sum
\be
\label{265}
 f_M \; = \; w_1 \; f_{M1} + w_2 \; f_{M2} \;   .
\ee
The Debye-Waller and Lamb-M\"{o}ssbauer factors for a phase $f$ enjoy the similar forms
\be
\label{266}
 f_{Mf} \; = \; \exp\left( - \; \frac{1}{3} \; k_0^2 \; r_f^2 \right) \;  .
\ee
Here $k_0$ is the momentum of a gamma-quantum in the case of the M\"{o}ssbauer effect and 
the momentum of an X-ray quantum or a neutron momentum in the case of the Debye-Waller 
factor. The quantity $r_f^2/3$ is the mean-square deviation in the direction of the 
X-ray quantum or gamma quantum, or neutron. 

The mean-square deviation for a phase $f$ in the Debye approximation is
\be
\label{267}
  r_f^2 \; = \; \frac{9 w_f \rho_f}{2m\Theta_f\rho} 
\int_0^1 x \; \coth\left( \frac{\Theta_f}{2T}\right) \; x \; dx \; ,
\ee
with the effective Debye temperature 
\be
\label{268}
 \Theta_f \; \equiv \; w_f^{3/2} \; T_{Df} \;  ,
\ee
where $T_{Df}$ is the Debye temperature of the pure phase $f$. At low and high temperatures, 
we have
$$
r_f^2 \; \simeq \; \frac{9\rho_f}{4m T_{Df}\sqrt{w_f} } \qquad ( T \ll T_{Df} ) \; ,
$$
\be
\label{269}
r_f^2 \; \simeq \; \frac{9\rho_f T}{m\rho T_{Df}^2 w_f^2 } \qquad ( T \gg T_{Df} ) \; .
\ee

The value of the factor $f_M$ essentially depends on phase probabilities, both explicitly
and through $f_{Mf} = f_{Mf}(w_f)$. The existence of mesoscopic fluctuations around the 
points of structural transitions can be characterized by the relative variation 
\be
\label{270}
 \dlt f_M \; \equiv \; \frac{f_M - f_{Mf}(1)}{f_{Mf}(1) } \;  .
\ee
To better stress the role of mesoscopic fluctuations, let us consider the case, when, 
in the absence of mesoscopic fluctuations, the factors $f_{Mf}(1)$ do not change much, 
so that $f_{M1}(1) \approx f_{M2}(1)$. And assume that, at the point of a phase 
transition, $w \approx 1/2$. Then the presence of mesoscopic fluctuations results in 
the sagging at the transition point of the factor $f_M$ described by the variation in 
the interval
\be
\label{271}
 f_{Mf}^3 -1 \; < \; \dlt f_M \; < \; f_{Mf}^{0.414} -1 \;  .
\ee
For instance, if $f_{Mf} = 0.9$, then $-0.27 < \delta f_M < -0.043$. In many cases, the 
transition temperature is higher then the Debye temperature, $T_c \gg T_{Df}$. In that 
case the sagging is estimated as $\delta f_M \approx -0.27$. This kind of saggings of 
Lamb-M\"{o}ssbauer factors have been observed at several structural phase transitions 
\cite{Bhide_63,Thosar_286}.

\section{Conclusion}

The problem of describing mesoscopic phase fluctuations in many-body systems is addressed.
A general approach is developed allowing for taking account of these fluctuations. Several
models of condensed-matter systems with mesoscopic fluctuations are considered. 

The basic idea of describing mesoscopic fluctuations is as follows. Suppose a statistical 
system is given, with microscopic states forming a Hilbert space $\mathcal{H}$. Generally, 
this space contains various microscopic states corresponding to different thermodynamic 
phases allowed for that system. The consideration of each phase requires an appropriate 
mathematical description satisfying the constraints characterizing the considered phase. 
This implies that not all microscopic states of the total Hilbert space are equally 
important for describing the particular chosen phase, since there are typical states 
characterizing different particular phases. In the simplest case, the definition of 
observable quantities for a particular phase can be done by employing the method of 
restricted trace selecting only those microscopic states that are typical of the studied 
phase. More generally, one can introduce the weighted Hilbert spaces $\mathcal{H}_f$, 
where the admissible microscopic states are weighted, with the higher weights ascribed 
to the considered phase $f$. In practice, the weighting can be realized by imposing 
appropriate constraints on the observable quantities. The direct sum of the weighted 
Hilbert spaces, $\bigoplus_f \mathcal{H}_f$, characterizes all admissible phases of the 
studied statistical system. All calculations for a particular phase $f$ are always 
accomplished in the related space $\mathcal{H}_f$. This is a kind of Hilbert space 
localization \cite{Cohen_287}. Thus the space $\bigoplus_f \mathcal{H}_f$ shows that 
the system can be in one of the phases, either phase $f=1$ or $f=2$, etc, but no 
coexistence of phases is possible. The description of several coexisting phases requires 
the use of the fibered Hilbert space $\bigotimes_f \mathcal{H}_f$. The definition of 
pure thermodynamic phases often needs to consider symmetries corresponding to the phases 
and the concept of symmetry breaking, which requires to resort to the thermodynamic limit. 
Nevertheless, sufficiently large finite systems allow for the definition of asymptotic 
symmetries \cite{Birman_288}, which makes it straightforward to deal with finite systems. 
The latter is important as far as mesoscopic fluctuations in finite systems can be even 
more pronounced, for example being due to the boundary effects.  

The theory of heterophase systems exhibiting mesoscopic phase fluctuations is illustrated
by several models including magnetic materials with paramagnetic fluctuations, 
ferroelectrics with paraelectric fluctuations, systems with mesoscopic density 
fluctuations, crystalline structures with structural fluctuations, frustrated materials
with coexisting crystalline and random structures, solids with superfluid fluctuations 
inside the regions of disorder, and superconductors with mesoscopic normal-phase 
fluctuations. The influence of mesoscopic fluctuations on the Debye-Waller and M\"{o}ssbauer 
factors is demonstrated.  

\vskip 5mm
{\bf Statements and Declarations}

\vskip 5mm
{\bf CONFLICT OF INTEREST}: The authors declare no conflicts of interests and no 
financial interests.

\vskip 5mm
{\bf AUTHOR CONTRIBUTIONS}: All authors contributed to the study conception and design. 
Material preparation, data collection and analysis were performed by V.I. Yukalov and 
E.P. Yukalova. The first draft of the manuscript was written by V.I. Yukalov and all 
authors commented on previous versions of the manuscript. All authors read and approved 
the final manuscript.

\vskip 5mm
{\bf FUNDING}: No funding was received to assist with the preparation of this 
manuscript.

\newpage
\begin{center}
{\Large {\bf Figure captions} }
\end{center}

\vskip 2cm
{\bf Figure 1}. Free energies as functions of the frustration parameter $g$ under the 
fixed cohesive energy parameter $u = -1$. 
(a) Free energy of the frustrated system $f$ (solid line), compared with the free 
energies of a pure crystalline solid $f_{sol}$ (dashed line) and a pure random matter 
$f_{ran}$ (dashed-dotted line); 
(b) Detalized region of the frustration parameter $g$, where the free energy of the 
frustrated matter crosses the free energy of the crystalline solid.

\vskip 1cm
{\bf Figure 2}. The probability of the crystalline $w_1$ (solid line) and random 
$w_2$ (dashed line) fractions in a frustrated material as functions of the frustration 
parameter $g$, under fixed $u = -1$. The point of the phase transition between the 
frustrated matter and crystalline solid is $g_0 = 0.128$; the upper bound for the 
existence of the frustrated matter is $g_c = 0.231632$.

\end{document}